\documentclass[aps,reprint,nofootinbib,longbibliography,onecolumn,prd]{revtex4-2}
\usepackage{bm}
\usepackage[latin1]{inputenc}
\usepackage[T1]{fontenc} 
\usepackage{graphicx,multirow}
\usepackage{float}
\usepackage{braket}
\usepackage{amsmath}
\usepackage{comment}
\usepackage{amssymb}
\usepackage{amscd}
\usepackage{latexsym}
\usepackage{slashed}
\usepackage{color, xcolor}
\usepackage{graphicx}
\usepackage[normalem]{ulem}
\definecolor{orange}{cmyk}{0,0.5,1,0}
\usepackage{verbatim}
\usepackage{rotating}
\usepackage{cancel}
\usepackage{hyperref}
\usepackage{lineno}
\usepackage{makecell}

\usepackage{makecell,natbib}
\hypersetup{
   colorlinks=true,       
   linkcolor=blue,        
   citecolor=red,         
   filecolor=magenta,      
   }

\begin{document} 

\title{Indirect detection of boosted light scalar dark matter}  

\author{Arindam Basu}
\affiliation{Department of Physics, School of Engineering and Sciences, SRM University-AP, Amaravati, Mangalagiri 522240, India}






\begin{abstract}
We explore the possibilities of detecting MeV-scale boosted dark matter (DM) via astrophysical observations. Given a particular model framework, using gamma-ray data from \texttt{COMPTEL} and \texttt{EGRET}, as well as neutrino constraints from the \texttt{Super-Kamiokande}, \texttt{Hyper-Kamiokande}, \texttt{JUNO}, diffuse supernova neutrino background (\texttt{DSNB}), and atmospheric neutrino observations, we find that boosted light scalar DM could be probed within the mass range of $\sim 20\!-\!80$ MeV, satisfying the Planck observed relic abundance. Our study shows, indirect search can play a key role in the detection of MeV-scale boosted DM, which are otherwise beyond the reach of usual collider or DM scattering experiments.
\end{abstract}


\maketitle

\setcounter{footnote}{0}




\section{Introduction}
\label{intro}
One of the leading hypotheses for the composition of dark matter is Cold Dark Matter (CDM), which consists of slow-moving, non-relativistic particles that played a critical role in the formation of large-scale structures in the Universe. As a key component of the Standard Cosmological Model, CDM provides a consistent explanation for a wide range of observations \cite{Planck:2018vyg,Bullock:1999he}, including galaxy rotation curves \cite{Zwicky:1933gu,Begeman:1991iy,Bertone:2010zza,Bauer:2017qwy,Bertone:2004pz,Lisanti:2016jxe}, gravitational lensing, and fluctuations in the Cosmic Microwave Background (CMB). The success of CDM in describing the distribution of galaxies and cosmic structures from the early Universe to the present strongly supports its validity. CDM is also essential for reproducing the observed large-scale structure, as it provides the necessary gravitational potential wells around which baryonic matter can clump and form galaxies. A well-motivated class of CDM candidates is Weakly Interacting Massive Particles (WIMPs) \cite{refId0,Arcadi:2017kky}. WIMPs naturally arise in several extensions of the Standard Model (SM) of particle physics and possess the right properties, such as being stable, electrically neutral, and interacting only weakly with ordinary matter to serve as viable CDM constituents. Their relic abundance, determined through thermal freeze-out in the early Universe, can naturally fall within the observed range of the DM density $(\Omega h^2 \sim 0.11-0.13)$ \cite{Planck:2015fie}, making WIMPs a compelling and extensively studied candidate for the particle nature of CDM. A plethora of search mechanisms for WIMP DM direct and indirect detection, with a very minimal success. The current scenario of the WIMP paradigm (In GeV scale) is discussed in \cite{Arcadi:2024ukq}, where most of the bidimensional plane $(m_{DM},\sigma_{DM}^{SI})$ in DM detection is searched with null results. Latest experiments like \texttt{XENONnT} \cite{XENON:2023cxc} and \texttt{LZ} \cite{LZ:2022lsv} have excluded spin-independent WIMP nucleon cross-sections down to the $\sim 10^{-47} - 10^{-48} \text{cm}^2$ level for WIMP masses around $20-40 \,\,\text{GeV}/\text{c}^2$, pushing sensitivity close to the so-called neutrino floor at lower masses, making it harder to probe.   

This pitfall of WIMP detection paves the way to explore the DM with low mass (typically MeV-scale). The issue with the MeV-scale DM candidate is the insufficient energy to recoil an electron/nucleon in the Earth-based detector medium. This imposes the condition on the light DM candidate that it should contain enough kinetic energy to overshoot the threshold recoil energy. Having extra kinetic energy, on top of the rest mass energy, is called a boost. The DM can be boosted via various sources; such as, cosmic ray boosting is the most studied domain of DM detection \cite{Bringmann:2018cvk,Das:2021lcr,Xia:2022tid,Jho:2021rmn,  Yin:2018yjn}, boost by the non-galactic sources \cite{Herrera:2023fpq,Herrera:2021puj}, and blazars \cite{Bhowmick:2022zkj,Wang:2021jic,Granelli:2022ysi,Maity:2022exk}. However, considering that light DM production occurs from a heavy DM's annihilation, makes the lighter DM boosted, due to the mass ratio. Several studies show this multi-component DM scenario \cite{Agashe:2015xkj,Aoki:2018gjf,Nagao:2024itk}. However, a particular model-based study is less discussed because it does not provide the freedom to play along with the parameters to satisfy the detection prospects. What it does provide is that a well-studied model, checked with theoretical and collider limits, intuitively addresses a DM, where the DM is first checked with the correct relic density. Thus, following a particular model, where the parameter space satisfies the various theoretical, experimental (collider) limits, the thermal evolution of the DM in this parameter space is well motivated.

The present paper is a subsequent analysis of our previous work, focusing on the detection prospects of the boosted MeV-scale scalar DM. In the previous paper \cite{PhysRevD.111.095016}, supported by various phenomenological constraints coming from different experiments such as the Large Electron-Positron Collider (LEP), Large Hadron Collider (LHC), etc \cite{Grimus:2007if,Grimus:2008nb,Haber:2010bw}, the model addresses the relic density aspects of the two-component boosted DM scenario. Where one DM candidate is a heavy (mass at $10\!-\!100$ GeV range) Majorana fermion, the other is light (mass at $10\!-\!100$ MeV range) scalar DM. The mass hierarchy of the two DM candidates, interacting through the Higgs portal, makes the lighter one boosted. We find a parameter space on the mass plane of the two DM candidates, satisfying the correct relic density. Following the allowed mass range, this present paper addresses the detection prospects of the boosted scalar MeV-scale DM. We show the scattering cross-sections in the present scenario are beyond the sensitivity reach of experiments, such as \texttt{XENON10} \cite{XENON10:2011prx}, \texttt{XENON1T} \cite{XENON:2019gfn}, \texttt{Darkside50} \cite{DarkSide:2022dhx}, and \texttt{SENSEI} experiment at Snolab \cite{SENSEI:2023zdf}. Moreover, due to the suppressed cross-section, it ends up inside the neutrino-floor, making it harder to detect. This motivates us to explore the fate of this model at the indirect detection frontier. As opposed to direct detection, indirect detection searches for photon or neutrino flux originating from DM annihilation/decay~\cite{Photino.55.257, Ullio:2002pj}. We bound the gamma-ray spectrum with the existing experiments such as $\texttt{COMPTEL}$ \cite{dissertation} and $\texttt{EGRET}$ \cite{Strong:2004de}, since the experiments provide the spectrum in the MeV range, aligned with our framework's energy scale. While for the neutrino spectrum, the diffuse supernova neutrino background (\texttt{DSNB}) \cite{Vitagliano:2019yzm}, and atmospheric neutrinos \cite{Peres:2009xe} provide the upper limit. In this energy scale, the experiments such as the \texttt{Super-Kamiokande} (\texttt{SK}) \cite{Super-Kamiokande:2002hei}, \texttt{JUNO} \cite{JUNO:2023vyz}, \texttt{Hyper-Kamiokande} (\texttt{HK}) \cite{Olivares-DelCampo:2018pdl} impose bounds on the cross-section for the light scalar DM annihilating to produce neutrinos. In the context of the present model set-up, we find, the photon or neutrino spectrum could either be probed at the present experimental facilities or could be detected with future observatories. 

This paper is organized as follows. In Sec.~\ref{model}, we briefly describe the crux of the model and the relic density aspects of the two-component boosted DM scenario. In Sec.~\ref{direct-detection}, we show the direct detection prospects for the boosted light scalar DM. We calculate the nuclear and electron recoil energies and the scattering cross-section, shown against the exclusion limits of various experiments. The indirect detection is followed up in Sec.~\ref{Indirect-detection}, where the gamma-ray and neutrino spectra are examined in detail. Finally, we conclude with Sec.~\ref{conclusion}.

\section{Two-component boosted DM framework}
\label{model}

We work within the neutrinophilic Two-Higgs Doublet Model ($\nu$2HDM) \cite{Machado:2015sha,Nomura:2017jxb}, extended by a real scalar singlet and vectorlike leptons (VLLs), which together give rise to a two-component dark matter (DM) scenario. A detailed construction of the model and constraints has been presented in our earlier work \cite{PhysRevD.111.095016}; here we summarize the salient features relevant for the current study. The model follows the gauge symmetry as, 
\begin{align}
SU(3)_C \times SU(2)_L \times U(1)_Y \times Z_2 \times Z_2^{\rm DM},
\end{align} \label{eq:gauge-sym}

with two Higgs doublets, $\Phi_1$ (SM-like) and $\Phi_2$ (neutrinophilic). All SM fermions couple to $\Phi_1$ (contains vacuum expectation value (VEV) at EW scale, $v_1=\langle 0| \Phi_1|0\rangle$), while right-handed neutrinos $N_{R_i}$ couple to $\Phi_2$, thereby generating tiny active neutrino masses via suppressed Yukawa couplings and a small VEV $v_2=\langle 0| \Phi_2|0\rangle$, consistent with a low-scale seesaw mechanism. The additional real singlet scalar $\phi_3$, stabilized by a discrete $Z_2^{\rm DM}$ symmetry, contributes as a light scalar DM candidate. Its interactions with the Higgs doublets are governed by scalar portal couplings, while the physical mass is
$$
m_{\phi_3}^2 = \mu_{\phi_3}^2 + \kappa_1 v_1^2 + \kappa_2 v_2^2 .
$$

Moreover, to address electroweak precision tension (notably in the $S$ parameter) and to provide a second DM component, we add an $SU(2)_L$ doublet $N$ and a singlet $\chi$, both odd under $Z_2^{\rm DM}$. Their mixing leads to two neutral eigenstates, of which the lighter one, $\chi_1$, plays the role of fermionic DM. The heavier mass state is taken to be decoupled ($\sim$TeV scale). All relevant fields with their corresponding charges are listed in Tab.~\ref{table:1}.
\begin{table}[H]
\centering
\begin{tabular}{ |c|c|c|c|c|c| } 
\hline
Particle Name & $SU(2)_L$ Charges &  $U(1)_Y$ Charges & $Z_2$  Charges & $Z_2^{\rm DM}$ Charges\\
\hline
 $\Phi_1$ & 2 & 1 &  1 & 1\\
\hline 
$\Phi_2$ &  2 & 1  & -1 & 1 \\ 
\hline
$\phi_3$ & 1 &  0  &  1 & -1 \\ 
\hline 
$N_{R_{i}}$ & 1 & 0  &  -1  & 1\\
\hline
 $N$ & 2 & -1  &  1  & -1\\
\hline 
$\chi$ &  1 & 0 & 1  & -1 \\ 
\hline
\end{tabular}
\caption{The BSM fields and their charge assignments.}
\label{table:1}
\end{table}

The  Lagrangian for the relevant interactions is mentioned in Eqs.~[10 - 12]  of~\cite{PhysRevD.111.095016}, where the phenomenologically important couplings can be found. The novel feature of this framework is the possibility of boosted scalar DM, as $\chi_1 \chi_1 \to \phi_3 \phi_3$ annihilation produces relativistic $\phi_3$ particles. This non-trivial interplay modifies the thermal history and relic density of both components. A detailed analysis of the scalar potential, neutrino mass generation, constraints (EWPT, Higgs searches, invisible $Z$ decays, etc.), and model-building aspects can be found in Ref.~\cite{PhysRevD.111.095016}. Finally, we use the coupled Boltzmann equations to show the thermal evolution of the two DM species. We find a range over the masses for the two DM candidates, consistent with various theoretical and experimental constraints. satisfying Planck observed relic density:
$$
m_{\chi_1} \simeq 30\!-\!65~{\rm GeV}, 
\qquad
m_{\phi_3} \simeq 20\!-\!80~{\rm MeV}.
$$
The boosted $\phi_3$ component yields distinct phenomenological signatures, motivating further study through direct and indirect detection probes. For the studies of the detection prospects of the boosted $\phi_3$, the values of the relevant couplings, satisfying the theoretical, experimental, and the relic density bound, are as Tab.~\ref{tab:couplings}.

\begin{table}[H]
    \centering
    \begin{tabular}{|c|c|c|c|c|c|c|}\hline
        $\lambda_{h\phi_3\phi_3}$ &  $\lambda_{H\phi_3\phi_3}$ & $\lambda_{\phi_3\phi_3HH}$ &
        $\lambda_{hN_R\nu_L}$ & $\lambda_{hN_R\nu_L}$ & $\lambda_{hf\bar{f}}$ & $\lambda_{Hf\bar{f}}$\\ [1ex] \hline
        2.47 & $2.26\times 10^{-5}$ & $10^{-3}$ & $-2.29\times 10^{-9}$ & $2.82\times 10^{-4}$ & $\frac{m_f}{v} \times c_\beta$ & $\frac{m_f}{v}\times s_\beta$ \\  [1ex]\hline
    \end{tabular}
    \caption{Important couplings for the detection prospects. Where the total VEV $v=\sqrt{v_1^2+v_2^2}$, and the mixing angle $s_\beta= \sin \beta \simeq \tan \beta \simeq 8.13 \times 10^{-6} $, and $c_\beta = \cos \beta \simeq 1$ for $v_1 =246$ GeV and $v_2 = 2$ MeV. Here $h$ is the SM-like Higgs (fixed at 125 GeV), and $H$ denotes the CP-even neutral MeV-scale BSM Higgs (fixed at 20 MeV).}
    \label{tab:couplings}
\end{table}
\section{Direct detection issue for MeV scale scalar DM}
In this section, we discuss the direct detection prospects of the boosted light scalar DM. Direct detection experiments rely on the scattering of the DM with a target particle on Earth, which is probed by measuring the recoil energy of the scattered target particle ~\cite{Witten}. The recoil energy is measured in terms of ionization, scintillation, or phonon excitation. There exists an upper limit of the kinetic energy of the scattered target particle, $k_{max} = 2 v^2 \mu_{\text{DM-target}}^2/m_{\text{target}}$, where $v$ is the velocity of the incoming DM, $\mu_{\text{DM-target}}$ is the reduced mass of the DM-target system. Experiments with nuclear targets \cite{LUX:2016ggv, PandaX-II:2017hlx, XENON:2018voc}, which are sensitive to recoil energy $\geq$ keV, lose sensitivity for DM masses lower than a GeV. However, if the incoming DM possesses extra kinetic energy, or in other words, if it is boosted, albeit a low mass DM, it can produce the recoil energy above the detector threshold. Therefore, we first discuss the nuclear recoil and further show the limits for electron recoil for the direct detection prospects of the boosted light scalar DM $\phi_3$, whose relic allowed mass range is mentioned above.  
\label{direct-detection}

\subsection{Nuclear recoil}
Here, we explore the interaction of a boosted scalar MeV-scale DM candidate $\phi_3$ with nucleons, specifically protons, in the context of our model. While we focus on proton (Hydrogen) targets for simplicity and because of their relevance in low-threshold experiments, the formalism can be extended to heavier nuclei as well \cite{Ema:2020ulo}. We consider the nuclei as point particles for the kinematical calculations, and then the form factor is introduced to incorporate the finite size effect as discussed in \cite{Ema:2020ulo}. We use the dipole form factor as,
\begin{equation}
    G(\textbf{q}^2)=\frac{1}{(1+\textbf{q}^2/\Lambda^2)^2}
\end{equation}
where $\textbf{q}$ is the momentum change and the $\Lambda$ is the reduced nucleon mass. For Hydrogen $\Lambda=770$ MeV as discussed in \cite{Weisenpacher:2000ip}. Therefore, for the scattering process $\phi_3\, n \to \phi_3\, n$, the cross-section reads as,
\begin{align}\label{eq:dsig-dt-N}
     \frac{d \sigma}{dt} = \frac{(4m_n^2-t) n_n^2 F_n^2(t)}{4\pi\lambda(s,m_{\phi_3}^2,m_n^2)}\Big[\frac{(\lambda_{h\phi_3\phi_3}~\lambda_{hn})^2}{(t-m_h^2)^2}+\frac{(\lambda_{H\phi_3\phi_3}~\lambda_{Hn})^2}{(t-m_H^2)^2}+\frac{2\lambda_{h\phi_3\phi_3}~\lambda_{hn}\lambda_{H\phi_3\phi_3}~\lambda_{Hn}}{(t-m_h^2)(t-m_H^2)}\Big], 
\end{align}
where the $\lambda_{h(H)n}$ is the standard model Higgs boson h (BSM Higgs boson H) coupling with the nucleon. It is given as,
\begin{equation}
    \lambda_{h(H)n} = \lambda_{h(H)q\Bar{q}} \Big(\frac{m_n}{m_u}f_u^n + \frac{m_n}{m_d} f_d^n\Big).
\end{equation}
The $\lambda_{h(H)q\Bar{q}}$ denotes the SM Higgs $h$ (BSM Higgs $H$) coupling with the quarks.
The $f_u^n = 1.99 \times 10^{-2}$ and $f_d^n = 4.31 \times 10^{-2}$ denote the contribution of a particular quark mass $m_q$ to the nucleon mass $m_n$, discussed in \cite{PhysRevLett.115.092301}. The $n_n$ denotes the atomic number of nucleons in the nucleus of the considered atom. $m_n$ is the nucleon mass. The $F_n(t)$ is the form factor.  For Hydrogen atom, $n_n=1$, $m_n= 938$ MeV (proton mass), and $F_n(\textbf{q}^2)=G(\textbf{q}^2)$. The $\lambda (s,m_{\phi_3}^2,m_n^2)=s^2+m_{\phi_3}^4+m_n^4 -2\,s\,m_{\phi_3}^2 -2\,s\,m_n^2 - 2\, m_{\phi_3}^2\,m_n^2$ is the usual Kallen function, where the $s=m_{\phi_3}^2+m_n^2+2\,E_{\phi_3}\,m_n$, and $t=2\,m_n~(m_n-E_n)$. The $E_{\phi_3}$ denotes the incoming energy of the boosted $\phi_3$, and the $E_n$ denotes the outgoing nucleus energy.  
Since the $\phi_3$ produced from annihilation of $\chi_1$, $(\chi_1\chi_1 \to \phi_3 \phi_3)$, where $\chi_1$ is non-relativistic, it's energy $E_{\phi_3}$ reads as,
 \begin{eqnarray}
  E_{\phi_3}=\gamma_{\phi_3} m_{\phi_3} = m_{\phi_3} \times \frac{m_{\chi_1}}{m_{\phi_3}}\Big[1+\frac{(v_{\chi_1}/c)^2}{2}\Big],  \label{eq:Ephi3}
\end{eqnarray}
where the non-relativistic velocity of $\chi_1$ is $v_{\chi_1}=$220 Km/s. It shows the $\phi_3$ emerges with an energy approximately equal to the mass of $\chi_1$.
Following the prescription of \cite{Agashe:2015xkj}, we analyze the scattering kinematics in the rest frame of the target nucleon, where the incoming and outgoing momenta of $\phi_3$ and the nucleon $n$ are defined as,
 \begin{align}
     \text{Incident}\,\, \phi_3: p_1=(E_{\phi_3},\vec{p}) \quad \quad  \text{Scattered}\,\, \phi_3: p_3=(E_{\phi_3}^\prime,\vec{p^\prime}) \\
\text{Incident}\,\, n: p_2 = (m_n,0)  \quad \quad  \text{Scattered}\,\, n: p_4 = (E_n,\vec{\textbf{q}}).
 \end{align}
 
In the scattering event, in the rest frame of the nucleon, the recoil energy of the nucleon is formulated in terms of the relevant masses and the incident energy as follows,
\begin{align}
     E_n(m_{\phi_3},m_{\chi_1}) = m_n \frac{(E_{\phi_3}+m_n)^2+ (m_{\phi_3}^2-E_{\phi_3}^2)\cos \theta}{(E_{\phi_3}+m_n)^2-E_{\phi_3}^2+m_{\phi_3}^2}, \label{eq:EN}     
\end{align}
At a scattering angle of $180^\circ$, where the incoming and outgoing $\phi_3$ are back-to-back, the recoil energy reaches its maximum value, given by,
\begin{equation}
     E_n^{max}(m_{\phi_3},m_{\chi_1}) = m_n \frac{(E_{\phi_3}+m_n)^2+E_{\phi_3}^2-m_{\phi_3}^2}{(E_{\phi_3}+m_n)^2-E_{\phi_3}^2+m_{\phi_3}^2}. \label{eq:ENmax} 
\end{equation}

\begin{figure}[h!]
    \centering    \includegraphics[width=0.6\linewidth]{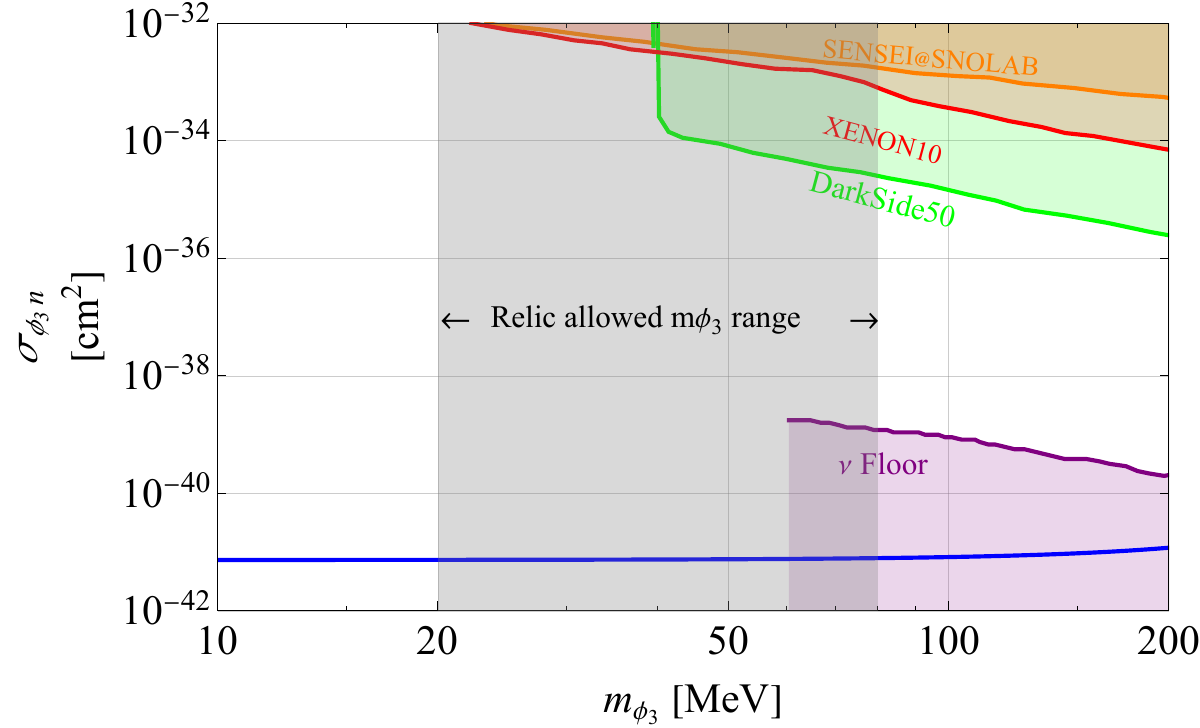}
    \caption{Boosted light scalar DM $\phi_3$-nucleon scattering cross-section shown by the blue curve. The exclusion limits for the MeV scale DM mass, shown by the green (\texttt{DarkSide}), red (\texttt{XENON10}), and orange (\texttt{SENSEI}) shaded regions. The neutrino floor is shown by the violet shaded region. The gray patch shows the allowed mass range for the correct relic density.}
    \label{fig:DM-N-sigma}
\end{figure}

To determine whether such a recoil is experimentally detectable, we evaluate the detector's threshold energy $E_n^{min}$, which is governed by the Cherenkov radiation threshold in the detection medium. In materials such as water or ice, the corresponding Lorentz boost factor is approximately $\gamma_{\rm Cherencov}=1.5$ \cite{Agashe:2015xkj}, meaning that the recoiling nucleon must acquire a kinetic energy exceeding this threshold to produce a detectable Cherenkov signal. Therefore, integrating the differential cross-section Eq. \ref{eq:dsig-dt-N} up to the maximum recoil energy $E_n^{max}$, we obtain the total cross-section $\sigma_{\phi_3 n}$ defined by the blue curve in Fig. \ref{fig:DM-N-sigma}. The form-factor dependency increases the cross-section for the higher mass regime.

Importantly, the model parameters follows Table.~\ref{tab:couplings}, which are consistent with the observed relic abundance. Due to the bound on the Higgs invisible branching ratio, the coupling $\lambda_{h\phi_3 \phi_3}$ is suppressed, while the coupling of $\phi_3$ and the nucleon with the BSM Higgs $H$ is suppressed by the smallness of the mixing angle $\beta$, demanding the alignment limit. 
Consequently, it results in a tiny cross-section as shown by the blue line in Fig.~\ref{fig:DM-N-sigma}. The model predicted line lies far below the experimental exclusion limits from \texttt{XENON10} \cite{XENON10:2011prx} (red shaded region), \texttt{SENSEI} \cite{SENSEI:2023zdf} (orange shaded region), and \texttt{DarkSide} \cite{DarkSide:2022dhx} (green shaded region). The \texttt{SENSEI} experiment at SNOLAB provides a strict limit on the DM-nucleon scattering cross section with light mediators, for dark matter masses below $\sim 40$ MeV. The $\texttt{XENON10}$ sets competitive bounds on the DM-nucleon scattering cross section, particularly for low mass dark matter and scenarios involving the Migdal-like processes~\cite{Essig:2019xkx}. 
The \texttt{DarkSide50} experiment at LNGS also uses the Migdal effect in liquid argon to extend sensitivity down to dark matter masses of $\sim 40$ MeV. It sets strong constraints on the spin-independent DM-nucleon scattering cross-section. Furthermore, a part of the calculated cross-section for the relic density allowed mass range: 20-80 MeV (shown by the gray shaded region) goes inside the neutrino floor ~\cite{Herrera:2023xun}, a region where coherent neutrino scattering backgrounds limit the sensitivity of direct detection experiments. 


In our analysis of nucleon scattering, we have adopted the dipole form factor. An alternative choice is the Helm form factor \cite{Duda:2006uk}, which accounts for the finite size of the nucleon. The Helm form factor modifies the uniform-sphere form factor by multiplying it with a Gaussian term, thereby incorporating the soft edge of the nucleus \cite{Helm:1956}. Due to its exponential dependence on the momentum change, the Helm form factor leads to an even stronger suppression of the scattering cross-section.
Therefore, we have not shown explicitly the bidimensional plane $(m_{\phi_3} - \sigma_{\phi_3 n})$ for the nucleon scattering considering the Helm form factor.
These results indicate that the traditionally dominant DM direct detection mechanism, the nuclear recoil, does not provide a viable detectable signal for this model, given the current and foreseeable experimental sensitivities.
This null result motivates exploring alternative scattering signatures, specifically electron recoil, which we discuss in the following section.
\subsection{Electron recoil}
\begin{figure}[h!]
    \centering
    \includegraphics[width=0.6\linewidth]{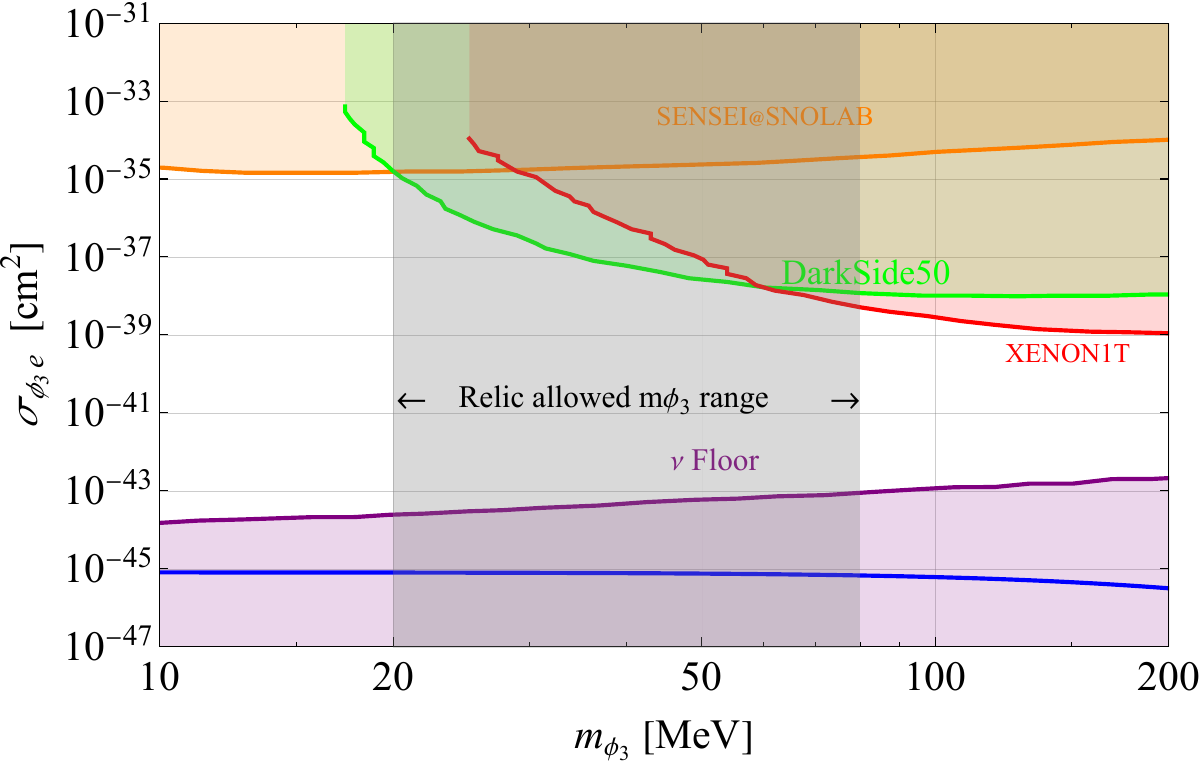}
    \caption{Boosted light scalar DM $\phi_3$-electron scattering cross-section shown by the blue curve. The exclusion limits for the MeV scale DM mass, shown by the green (\texttt{DarkSide}), red (\texttt{XENON1T}), and orange (\texttt{SENSEI}) shaded regions. The neutrino floor is shown by the violet shaded region. The gray patch shows the allowed mass range for the correct relic density.}
    \label{fig:DM-e-sigma}
\end{figure}

Electron recoil is an especially valuable channel for detecting light DM, particularly in the sub-GeV regime. In this mass range, DM typically lack sufficient kinetic energy to induce observable nuclear recoils, making electron scattering a more favorable detection strategy. This is due to the fact that electrons, being much lighter than nuclei, can receive larger recoil energies for the same energy of the incoming DM. Thus, DM-electron interactions offer a complementary window to probe light DM, and are being actively pursued in several low-threshold experiments like \texttt{SENSEI}, \texttt{XENON}-based detectors. In our setup, we consider the elastic scattering process.
\begin{align}
    \phi_3 e^- \to \phi_3 e^-,
 \end{align}
where $\phi_3$ is boosted. The kinematics to obtain the maximum electron recoil energy follows the same formalism as discussed above in the nuclear recoil section. It reads as,
\begin{equation}
    E_e^{max}(m_{\phi_3},m_{\chi_1}) = m_e \frac{(E_{\phi_3}+m_e)^2+E_{\phi_3}^2-m_{\phi_3}^2}{(E_{\phi_3}+m_e)^2-E_{\phi_3}^2+m_{\phi_3}^2}. \label{eq:Eemax}
\end{equation}
Here, $E_{\phi_3}$ is the energy of the incoming boosted $\phi_3$ as given in Eq. \ref{eq:Ephi3}. Following $E_{\phi_3} \simeq m_{\chi_1}$, the maximum recoil energy is of the order of $\chi_1$ mass, $E_e^{max} \simeq \mathcal{O}(m_{\chi_1})$. To determine whether such a recoil could be experimentally observed, we consider the detector threshold energy $E_e^{min}$, which depends on the Cherenkov radiation threshold in the detection medium.
\begin{align}
    E_e^{min} > \gamma_{\rm Cherenkov} \times m_e,
\end{align}
where the Lorentz boost factor is approximately $\gamma_{\rm Cherenocov}=1.5$. This implies that the minimum detectable recoil energy is relatively low. Given that the maximum recoil energy is of the order of $\chi_1$, condition $E_e^{max} >E_e^{min}$ is always satisfied for realistic parameter choices, thus ensuring a viable phase space for detection.
The $\phi_3 \!-\!e$ scattering cross-section can be conventionally normalized with a reference cross-section $\overline{\sigma_e}$ defined as \cite{Lin:2019uvt,Barman:2024lxy},
 \begin{align}
     \overline{\sigma_e} = \frac{\mu_{\phi_3 e}^2}{16\pi\, m_{\phi_3}^2\, m_e^2} |\overline{\mathcal{M}_{\phi_3 e}(|\vec{\textbf{q}}|)}|^2_{\textbf{q}=\alpha m_e},
 \end{align}
where the $\mu_{\phi_3 e}=\frac{m_{\phi_3}\,m_e}{m_{\phi_3}+m_e}$ is the reduced mass of the DM-e system, with the 3-momentum transfer $\textbf{q}$ fixed at the reference value $\alpha\,m_e$, which is the minimum required momentum to knock off an electron from its orbit. This comes with a criterion of $m_{\text{DM}}>m_e$, which is true in our case.
For the scattering process in concern $\phi_3 e^- \to \phi_3 e^-$ via a t-channel mediated by SM Higgs $h$, and the BSM Higgs $H$,  the square of the matrix element follows,
\begin{align}
    |\mathcal{M}_{\phi_3e}(\textbf{q})|^2 = (4m_e^2 + \textbf{q}^2)\Big[\frac{(\lambda_{h\phi_3\phi_3}~\lambda_{hee})^2}{(\textbf{q}^2+m_h^2)^2}+\frac{(\lambda_{H\phi_3\phi_3}~\lambda_{Hee})^2}{(\textbf{q}^2+m_H^2)^2}+\frac{2\lambda_{h\phi_3\phi_3}~\lambda_{hee}\lambda_{H\phi_3\phi_3}~\lambda_{Hee}}{(\textbf{q}^2+m_h^2)(\textbf{q}^2+m_H^2)}\Big],
\end{align}
where the square of the 3-momentum transfer $\textbf{q}$ is defined as, $|\textbf{q}|^2 =2m_e(m_e-E_e)$, as a function of electron-recoil energy $E_e$. For the realistic cross-section, the momentum dependence of the amplitude is captured in the form factor as  \cite{Lin:2019uvt,Barman:2024lxy},
\begin{align}
    F_{\phi_3e}(\textbf{q}) = \frac{|\mathcal{M}_{\phi_3e}(\textbf{q})|}{|\mathcal{M}_{\phi_3e}(\alpha\,m_e)|} .
\end{align}
Thus, the total cross-section can be expressed as \cite{Lin:2019uvt},
\begin{align}
    \sigma_{\phi_3 e} = \overline{\sigma_{e}} \times\int^{E_e^{max}} \frac{dE_e}{E_e^{max}} |F_{\phi_3e}(\textbf{q})|^2. \label{eq:dsig-dt-e}
\end{align}
In the non-relativistic case, the form-factor reduces to \cite{Barman:2024lxy},
\begin{align}
    F_{\phi_3e}(\textbf{q}) = \frac{m_H^2 + \alpha m_e^2}{m_H^2+\textbf{q}^2} \begin{cases}
        1 & m_H \gg \alpha\, m_e\,, \\ 
        \frac{\alpha^2 m_e^2}{\textbf{q}^2} & m_H \ll \alpha\,m_e\, . 
    \end{cases}
\end{align}
Therefore, given $\alpha\,m_e \simeq 3.7\times 10^{-6}$ GeV, $m_H\gg \alpha\,m_e$ limit is always valid, assuring $F_{\phi_3 e}=1$. However, the $\phi_3$ here is boosted, and thus we should refrain from assuming the non-relativistic approximation for the form-factor. Therefore, from the Eq. \ref{eq:dsig-dt-e}, integrating the form-factor up to the maximum electron recoil energy Eq.~\ref{eq:Eemax}, we get the blue solid line in Fig. \ref{fig:DM-e-sigma}. This line represents our model prediction for DM-electron scattering via the boosted $\phi_3$ component. Since, $\sigma_{\phi_3 e} \propto 1/m_{\phi_3}^2$, we see the falling nature of the blue curve at larger masses.
From this figure, it is evident that the predicted cross-section lies far below the exclusion limits set by direct detection searches such as \texttt{SENSEI} \cite{SENSEI:2023zdf}, \texttt{DarkSide50} \cite{DarkSide:2018bpj}, and \texttt{XENON1T} \cite{XENON:2019gfn}. In fact, the predicted signal reaches down into the neutrino floor \cite{PhysRevD.109.083016}, a region where solar neutrino backgrounds become dominant and make further direct detection highly challenging. This implies that, while theoretically consistent, the particular model would not yield observable signals in present-day detectors through the electron recoil channel as well. 

To bring the predicted signal into an experimentally accessible regime, one could, in principle, enhance the cross-section by increasing the DM-mediator (in our case, $\lambda_{h(H)\phi_3 \phi_3}$) and/or mediator-electron/nucleon (in our case, $\lambda_{h(H)ee}$\ $\lambda_{h(H)n}$) couplings. Such as in the traditional electrophilic model \cite{Trickle:2020oki}, due to the dominant coupling of a dark photon with the DM and the electron, one can probe the electron-positron excess observed by DAMPE, as well as obtain testable signatures in direct detection experiments via kinetic mixing with a dark photon \cite{Gu:2017gle}. Similarly, effective models with a hidden $U(1)_D$ gauge symmetry, which provides the dark photon, can address viable event rates for the MeV scale DM direct detection \cite{Essig:2011nj}. For example, Ref. \cite{Nagao:2024hit} demonstrates light DM detection in a similar framework to ours; a two-component boosted DM scenario, showing viable direct detection event rates for the boosted DM candidate in the effective parameter space of $(\lambda_{\text{dark\,\,photon - SM}})-\delta m$, where $\delta m$ is the mass difference between the two DM components in a quasi-degenerate case.
However, in our specific model-dependent scenario, we do not have the freedom to vary the mass splitting between the two DM candidates arbitrarily, since it is fixed within an allowed range to ensure the correct relic density. Furthermore, the couplings in our model are subject to multiple theoretical and experimental constraints such as perturbativity, unitarity, and limits from collider experiments discussed in Sec. \ref{model} and in our earlier study \cite{PhysRevD.111.095016}. In addition, it is important to note here that the scattering cross-section of the $\phi_3$ with the nucleon/electron as given in Eq.~\ref{eq:dsig-dt-N} and Eq.~\ref{eq:dsig-dt-e}, the dominant contribution arises from the $t$-channel exchange of the Higgs $h$. In the alignment limit, $h$ behaves as the SM Higgs, and therefore its couplings to SM particles are fixed to their SM values (with $m_h = 125$ GeV). This leaves no freedom to enhance these interactions. The coupling $\lambda_{h\phi_3\phi_3}$ is already taken at its maximum value allowed by the Higgs invisible decay bound, while other parameters do not play a significant role. Consequently, enhancing the cross-section beyond the neutrino floor, abiding all other bounds, remains difficult within this framework. 

To summarize, while the electron recoil channel offers a theoretically motivated and experimentally clean avenue for detecting light DM, our model, in its current form, remains in tension with experimental sensitivity. Future improvements in detector thresholds or the development of new search strategies targeting boosted scalar signatures may eventually bring such scenarios within reach. However, DM-nucleon/electron scattering cross-sections below the neutrino floor can be probed by directional detection experiments, with minimal sensitivity loss from the neutrino background, where the direction of the DM and neutrino is also measured along with the recoil energies. This exercise is beyond the scope of this paper. The DM directional detection strategies are discussed in~\cite{OHare:2015utx}. Given the insufficient sensitivity of both electron and nuclear recoil channels in direct detection, we are thus motivated to explore the prospects of indirect detection for this model.

\section{Indirect detection}
\label{Indirect-detection}

Unlike direct detection, which relies on detecting the recoil energy of the scattered target particle inside an Earth-based detector, indirect detection accounts for the fluxes of cosmic rays, neutrinos, and gamma rays originating from the annihilation/decay of DM, from the regions of the DM halo. Among these, the gamma-ray search and the neutrino spectrum study are promising due to their comparatively uninterrupted travel through space and the relatively low astrophysical backgrounds, making them a clean observational window into the DM indirect detection searches \cite{Bergstrom:1997fj}.
In addition, except for the kinematically forbidden annihilation/decay channels of light DM, another potential indirect detection signal for light DM comes from the electron-positron ($e^\pm$) spectrum \cite{Coogan:2019uij, Arina:2025ner}. However, in our present model, the branching ratio (BR) of $H$ into an $e^+e^-$ pair is much smaller than its BR into neutrino or photon pairs, making this channel subdominant.
In what follows, we explore the indirect detection prospects of our model, focusing on the gamma-ray and neutrino spectrum arising from the annihilation or decay channels of the boosted light scalar DM candidate. 
\subsection{Gamma ray spectral features from cascaded annihilations}
\label{Gamma-ray}

\begin{figure}[h!]
    \centering
    \includegraphics[scale=0.35]{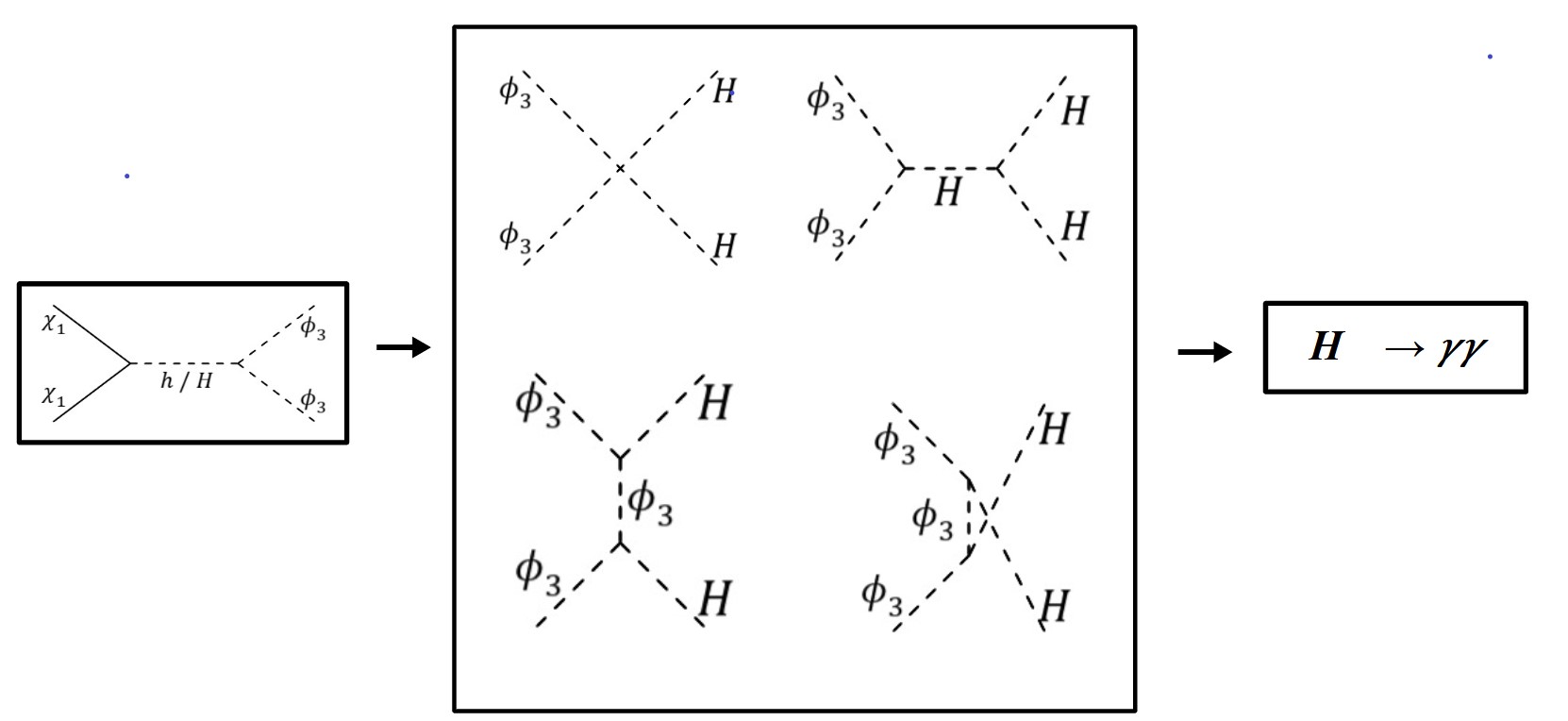}
    \caption{Schematic diagram of the process for the $\phi_3 \,\phi_3 \to \gamma\,\gamma\,\gamma\,\gamma$}
    \label{fig:diagram}
\end{figure}

Due to the production of $\phi_3$ via 2-to-2 annihilation of  $\chi_1$, the total center of mass (COM) energy of the incoming $\phi_3$'s become,
\begin{align}\label{eq:boost-COM}
    S= 4m_{\phi_3}^2 + 4m_{\phi_3}^2 \Big(1- \frac{m_{\phi_3}^2}{m_{\chi_1}^2} (1-v_{\chi_1}^2)\Big),
\end{align}
where the $v_{\chi_1}=220$ km/s is the non-relativistic velocity of the parent DM candidate $\chi_1$ inside the DM halo. As discussed earlier, to satisfy the correct relic density in this particular two-component scenario, the maximum allowed mass for the fermionic DM is 65 GeV, resulting in the COM energy of $S \simeq 8m_{\phi_3}^2$. Thus, for the second step, where $\phi_3$ annihilates to the pair of $H$, the total initial energy of the two incoming $\phi_3$'s is $2\sqrt{2}~ m_{\phi_3}$, making each $H$ to contain $\sqrt{2}~m_{\phi_3}$ energy. Now, each of these $H$ decays into a pair of photons, which in the rest frame of $H$ carries energy $m_H/2$. Thus, in the lab frame, the photon energy reads,
\begin{align} \label{eq:photon-energy}
    E_{\gamma} = \frac{m_{\phi_3}}{\sqrt{2}} \Big( 1+ \cos \theta \sqrt{1-\frac{m_H^2}{2\,m_{\phi_3}^2}}~ \Big)\,,
\end{align}
where $\theta$ is the angle between the Higgs $H$ and the outgoing photons. The detail of spectrum kinematics is given in Sec.~\ref{spectrum-energy}.
Since the decaying particle is a scalar, the photon emission is isotropic. Therefore, the resulting spectrum takes a box-shaped form between the two end points \cite{Ibarra:2012dw, Ibarra:2013eda, Boddy:2015efa},
\begin{align} \label{eq:photon-energy-spectra}
    \frac{dN_{\gamma}}{dE_{\gamma}} = \frac{4}{\big(E_{\gamma}^{max} - E_{\gamma}^{min}\big)} \Theta\big(E-E_{\gamma}^{min}\big) \Theta \big(E_{\gamma}^{max} - E\big),
\end{align}
where $E_{\gamma}^{max}$ and $E_{\gamma}^{min}$ are related to the maximum and minimum energy of a photon for $\theta = 0^\circ$ and $\theta= 180^\circ$ respectively, gives the edges of the box. The $E_c= \frac{E_{\gamma}^{max} +E_{\gamma}^{min}}{2} = \frac{m_{\phi_3}}{\sqrt{2}}$, while the width of the box is given by $\Delta E = E_{\gamma}^{max} - E_{\gamma}^{min} = 2 \frac{m_{\phi_3}}{\sqrt{2}} \sqrt{1-\frac{m_H^2}{2\,m_{\phi_3}^2}}$. It shows, for $\frac{m_H^2}{2\, m_{\phi_3}^2} \to 1$, the spectrum $\frac{dN_{\gamma}}{dE_{\gamma}}$ reduces to a line, while for higher DM masses, the width gets wider. 
Now, the flux of the photons induced by the annihilation reads,
\begin{align} \label{eq:gamma-flux1}
    \frac{d\widetilde{\phi_{\gamma}}}{dE_{\gamma}} = \frac{\langle \sigma v\rangle}{8\pi m_{\phi_3}^2} \frac{dN_{\gamma}}{dE_{\gamma}} \frac{1}{\Delta \Omega} \int_{\Delta \Omega} d \Omega \,J_{ann}\,,
\end{align}
where $\langle \sigma v\rangle$ is the thermal averaged cross-section of the boosted $\phi_3$'s self annihilation into pair of $H$. The cross-section expression is given in Eq.~\ref{eq:A1}.
Among the diagrams as shown in Fig.~\ref{fig:diagram}, the dominant contribution arises from the contact interaction with the quartic coupling $\lambda_{\phi_3\phi_3 HH}$, which sets the scale of both the annihilation rate and the flux.
Here, $\Delta \Omega$ is the observed field of view, and the $J_{ann} = \int_{l.o.s} ds\, \rho_{\phi_3}^2 $ is the integral of the squared DM density over the line of sight (l.o.s). For simplicity, considering the Navarro--Frenk--White (NFW) profile~\cite{Navarro:1995iw,Navarro:1996gj} for the DM density at the region of interest; the galactic center and halo regions as discussed in \cite{Vertongen:2011mu}, we use the features $\Delta \Omega = 1.30~ \text{sr}$, and $\int_{\Delta \Omega} d \Omega \,J_{ann} = 9.2 \times 10^{22} ~\text{GeV}^2\, \text{cm}^{-5} \,\text{sr}$. In spectral analysis, it is necessary to convolve the expected flux $\widetilde{\phi_{\gamma}}$ with the experimental energy resolution. Assuming a Gaussian detection response characterised by a standard deviation $\sigma(E)$, the convoluted photon flux reads,
\begin{align} \label{eq:gamma-flux2}
    \frac{d\phi_{\gamma}}{dE_{\gamma}} = \int dE \, \frac{d\widetilde{\phi_{\gamma}}}{dE} \frac{1}{\sqrt{2\pi}\, \sigma(E)} \text{exp}\Big( -\frac{(E-E_{\gamma})^2}{2 \sigma^2(E)} \Big)\,.
\end{align}
The energy resolution of the COMPTEL experiment is a maximum of 10\% \cite{dissertation}. This rounds off the sharp edges present in the $\widetilde{\phi_{\gamma}}$. 
\begin{figure}[h!]
    \centering
    \includegraphics[scale=0.45]{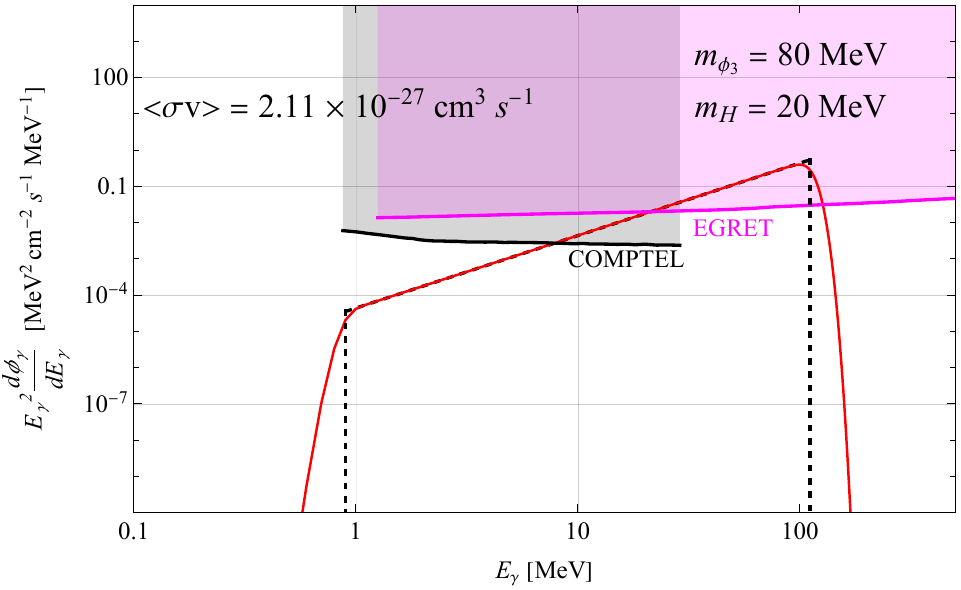} \quad \includegraphics[scale=0.45]{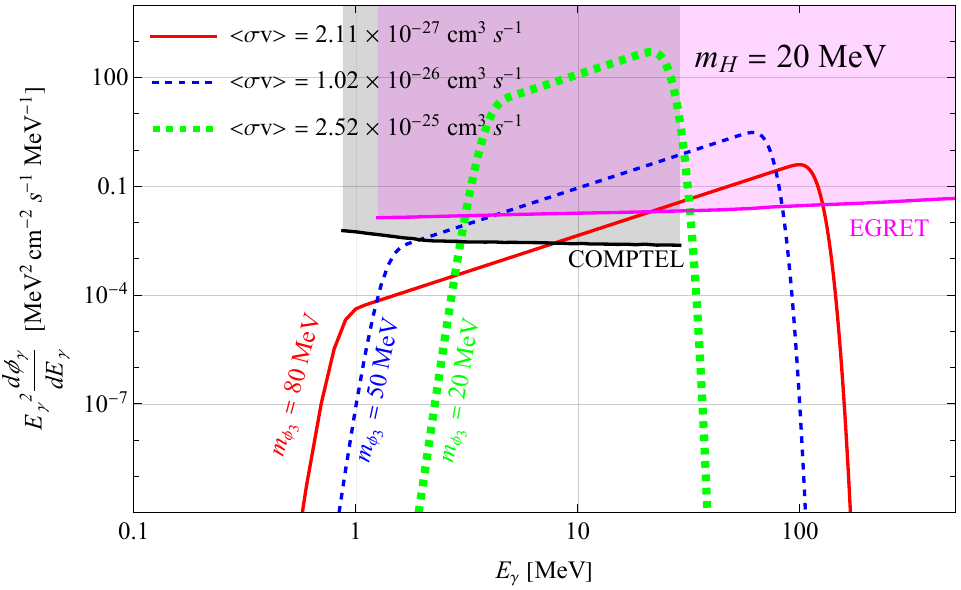}
    \caption{The spectrum of box-shaped gamma rays. The left panel shows the flux (black dashed) and convoluted flux (red solid) for the $m_{\phi_3} = 80$ MeV, and $m_{H}= 20$ MeV. The thermal averaged cross-section is $\langle \sigma v\rangle = 2.11 \times 10^{-27} \text{cm}^3 \text{s}^{-1}$. The right panel shows the convoluted flux for $m_{\phi_3}=$ 80 MeV (red solid), 50 MeV (blue dashed), and 20 MeV (green dotted), for fixed $m_H=20$ MeV. For each choice, the thermal averaged cross-section is marked inside. The sensitivity lines from the COMPTEL and EGRET data are shown in blue and green solid lines, respectively.}
    \label{fig:boxy}
\end{figure}
The left panel of Fig. \ref{fig:boxy} shows the convoluted and unconvoluted flux by the red solid and black dashed curves, respectively. The masses and the thermal averaged cross-section are mentioned in the plot legend. The right panel shows the convoluted flux for different choices of DM masses within the relic allowed region for the scalar DM mass $m_{\phi_3}$. The $E_c$ and width of the box-shaped spectrum depend on $E_{\gamma}^{max}$ and $E_{\gamma}^{min}$. However, due to the log scale, it appears that the spectra for different mass choices center around 10 MeV. Therefore, we show a quantitative measurement of the $E_{\gamma}^{max}$, $E_{\gamma}^{min}$, and $E_c$ in Tab. \ref{tab:energy}.
\begin{table}[h!]
    \centering
    \begin{tabular}{|c|c|c|c|} \hline
        $m_{\phi_3}$ (MeV) & $E_{\gamma}^{max}$ (MeV) & $E_{\gamma}^{min}$ (MeV) & $E_c$ (MeV) \\ \hline 
        80 & 112 & 0.8 & 56 \\ 
        50 & 69 & 1.4 & 35 \\
        20 & 24 & 4.1 & 14 \\ \hline
    \end{tabular}
    \caption{Maximum, minimum, and $E_c$ values of the photon spectrum energy, for different choices of DM mass which satisfy the correct relic density.}
    \label{tab:energy}
\end{table}
The black and magenta shaded regions show the excluded region of the parameter space due to the upper limit on flux measured by \texttt{COMPTEL}~\cite{dissertation} and \texttt{EGRET}~\cite{Strong:2004de}, respectively. \texttt{COMPTEL} provides measurements of large-scale gamma ray intensity in the MeV regime, particularly $0.8 - 30$ MeV, focusing on the full-sky diffuse component after removal of resolved point sources and instrumental backgrounds. The $\texttt{EGRET}$ operating on $1 ~\text{MeV} - 10 ~\text{GeV}$ mapped the diffuse Galactic and extragalactic gamma-ray emission, using detailed modeling of the Galactic plane and mid-latitude regions (e.g $5^\circ \leq |b| \leq 36^\circ$) to derive intensity limits for the diffuse background. 
The right panel shows that for the mass choice $m_{\phi_3} =$ 80 MeV the portion of the plateau region corresponding to higher gamma-ray energies lies above the experimental exclusion bounds, placing this mass choice and the associated energy range in tension with experimental data. In contrast, at lower $E_\gamma$, the same mass choice remains valid within the observable sensitivity range. For $m_{\phi_3}=$50 MeV, the predicted flux lies partially within the excluded region, while for $m_{\phi_3}=$20 MeV it falls predominantly inside the exclusion bounds, rendering these cases less promising for detection. Future improvements in sensitivity, such as those expected from missions like $\texttt{e-ASTROGAM}$ and $\texttt{AMEGO}$ \cite{Bartels:2017dpb}, could provide more stringent bounds on the flux and either validate or further constrain the parameter space.
It is worth noting that the limits derived here are consistent with the annihilation cross-section bounds discussed in \cite{Cirelli:2025rky}, where the upper limit of $\langle \sigma v \rangle$ is found to be of the order of $10^{-27} \text{cm}^3/\text{s}$, for similar mass ranges. In our analysis, we also find that scenarios with $\langle \sigma v \rangle  \lesssim 10^{-27} \text{cm}^3/\text{s}$ remain within the sensitivity reach, while any significantly larger cross-section will be in tension with gamma-ray search experiments. Relatively relaxed bounds are explored in~\cite{Guo:2023kqt}, where an allowed band of $10^{-28} \lesssim \langle \sigma v \rangle \lesssim 10^{-22} \text{cm}^3/\text{s}$ is obtained by combining different sets of gamma-ray experimental data and varying the DM-mediator mass combination. In that scenario, the DM predominantly undergoes semi-annihilation to produce a complex scalar mediator, which subsequently decays into a photon pair, leading to distinctive gamma-ray signatures, which is forbidden by the underlying symmetry for the present model [cf. Eq.~\ref{eq:gauge-sym}].
\subsection{Gamma ray spectral features due to prompt annihilation}
\label{2gamma-final-state}
\begin{figure}[h!]
    \centering
    \includegraphics[scale=0.45]{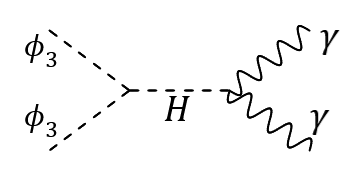}
    \caption{Feynamann diagram for $\phi_3\,\phi_3\to \gamma\,\gamma$, the prompt photon production from the self-annihilation of $\phi_3$ with $H$ mediation.}   
    \label{fig:2gamma}
\end{figure}

\begin{figure}[h!]
    \centering
    \includegraphics[scale=0.45]{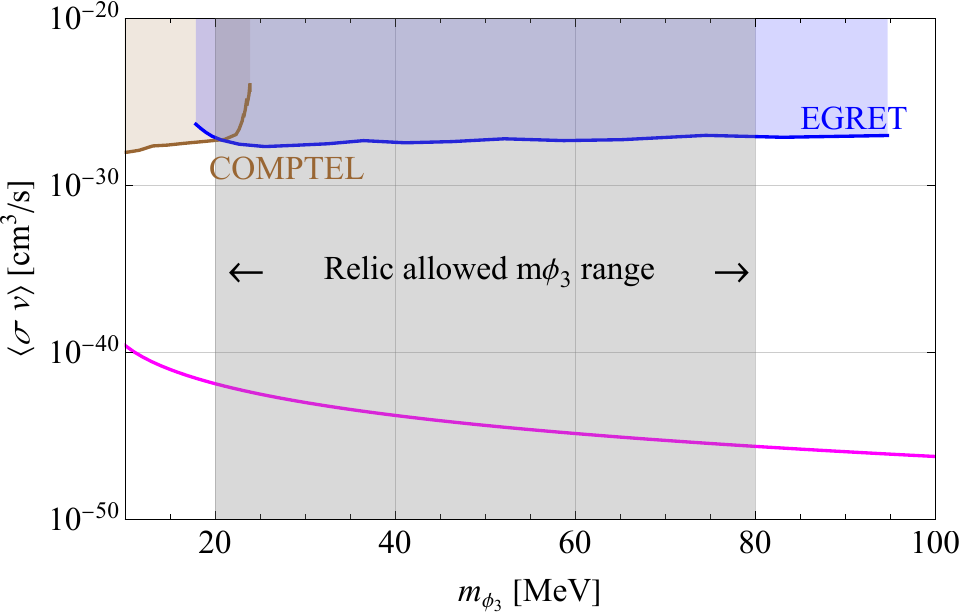} \quad \includegraphics[scale=0.45]{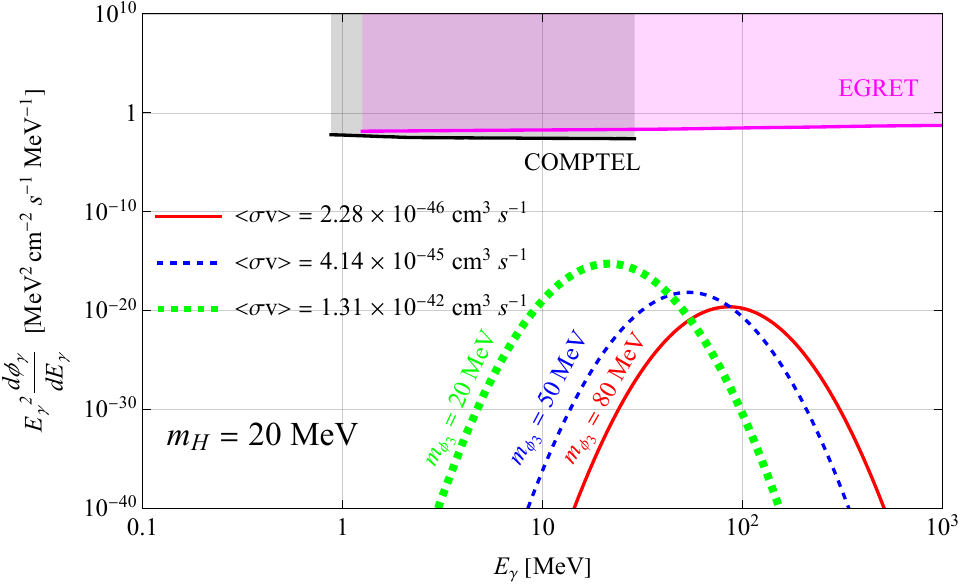}
    \caption{Left: exclusion limits on the thermal-averaged cross-section for the $\phi_3\,\phi_3 \to \gamma \,\gamma$ process in our galaxy from \texttt{COMPTEL} (brown shaded), and \texttt{EGRET} (blue shaded) sensitivity. The gray shaded region shows the allowed mass range for the correct relic density. Right: the flux of the gamma spectrum generated due to the photon excess produced directly from the DM self-annihilation. The monochromatic gamma-ray flux is shown for the boosted scalar DM $\phi_3$ mass 80 MeV (red solid line), 50 MeV (blue dashed line), and 20 MeV (green dotted line). The bounds are from \texttt{COMPTEL}, and \texttt{EGRET} is shown by the black and magenta shaded regions.}
    \label{fig:2gamma-spectra}
\end{figure}

In parallel to the cascaded annihilation of $\phi_3$, which gives rise to the broad gamma-ray spectrum, it can also undergo direct self-annihilation into a pair of photons. This process is illustrated in Fig.~\ref{fig:2gamma}, which provides a monochromatic photon signal, often considered a smoking-gun signature for DM annihilation. Here, we focus on the $H$ mediated annihilation. The alternative possibility of SM-like Higgs $h$ mediation is kinematically suppressed and thus negligible in our parameter space. The corresponding cross-section for $\phi_3\,\phi_3 \to \gamma\,\gamma$ is mentioned in Eq.~\ref{eq:A2}, while its thermal average is displayed in the left panel of Fig.~\ref{fig:2gamma-spectra} (solid magenta line). As seen from the figure, the thermal averaged cross-section decreases with increasing DM mass, which is due to the scaling $\sigma_{\phi_3\phi_3 \to \gamma\gamma} \propto 1/m_{\phi_3}^2$. The gray shaded region indicates the mass window consistent with the observed relic density. Importantly, in this region, the predicted annihilation cross-section lies well below the upper limits provided by \texttt{COMPTEL} (brown solid line) and \texttt{EGRET} (blue solid line).
The brown and blue shaded regions indicate the parameter space excluded by these experiments, based on data compiled in \cite{Boddy:2015efa}. 
The cross-section is controlled by the coupling strength of the mediator $H$ with the $\phi_3$ and the photon. These interactions are governed by the parameters $\lambda_{H\phi_3\phi_3}$ and $\lambda_{Hf\bar{f}}$ (which feed into the decay width $\Gamma_{H\to \gamma\gamma}$ \cite{Djouadi:2005gi}). Due to small mixing ($\tan \beta = 8.13 \times 10^{-6}$ see ref. \cite{PhysRevD.111.095016}), the cross-section is suppressed. Interestingly, while such small couplings diminish direct-detection prospects, as discussed before, they turn out to be advantageous here. The predicted signal, albeit remains beyond the present exclusion bound, yet stays within the sensitivity reach of future gamma-ray telescopes. Thus, the two-photon final state provides a promising probe of this scenario.

The spectra of the prompt photons produced in the process $\phi_3\phi_3 \to \gamma\gamma$, show that the resulting energy distribution is sharply peaked. In the rest frame of the annihilating DM, each photon carries away energy same as the mass of $\phi_3$, and hence the spectrum is simply described by a $\delta$-function,
\begin{align}
    \frac{dN_\gamma}{dE_\gamma} = \delta(E_\gamma - m_{\phi_3})\,.
\end{align}
The above equation reflects the fact that the signal corresponds to a monochromatic line at energy $E_\gamma = m_{\phi_3}$. To connect this idealized spectrum with realistic observations, one must account for the finite energy resolution of detectors. Following the same flux as in Eq. \ref{eq:gamma-flux1}, the primary modification here is the convolution with an appropriate resolution function. Since the flux peaks at a single energy, the measured spectrum is obtained by smearing the $\delta$-function peak around $E_\gamma = m_{\phi_3}$ using the instrumental response. We adopt a Gaussian energy resolution with width $\sigma = 10\%$ \cite{dissertation}, centered around the photon line \cite{Palomares-Ruiz:2007egs}. The observed flux is therefore given by,
\begin{align} \label{eq:gamma-flux3}
    \frac{d\phi_{\gamma}}{dE_{\gamma}} = \int_0^\infty dE \, \frac{d\widetilde{\phi_{\gamma}}}{dE} \mathcal{G}(E,E_\gamma)\,,
\end{align}
where the convolution function $\mathcal{G}(E,E_\gamma)$ encodes the detector resolution and reads \cite{Akita:2025dhg} as,
\begin{align} \label{eq:Convolv-func}
    \mathcal{G}(E,E_\gamma) = \frac{e^{-\frac{1}{2}(\sigma \,\text{ln}\,10)^2}}{\sqrt{2\pi\,\sigma} \,\text{ln}\,10}\, \frac{1}{E} e^{-\frac{1}{2}\Big( \frac{\text{log}_{10}\, E_\gamma/E}{\sigma}\Big)^2}.
\end{align}

The right panel of Fig. \ref{fig:2gamma-spectra} displays the gamma-ray flux from prompt photons for three benchmark $\phi_3$ masses: $m_{\phi_3} = 80$ MeV (red solid line), 50 MeV (blue dashed line), and 20 MeV (green dotted line), that satisfy the observed DM abundance. The spectra are shown in the plane of $E_\gamma^2 \, \frac{d\phi_\gamma}{dE_\gamma}$ versus photon energy $E_\gamma$. As the mass increases, the peak of the flux decreases, which directly follows from the reduction in the annihilation cross-section, as the left panel of Fig. \ref{fig:2gamma-spectra} delineates. Also note that increasing $m_{\phi_3}$ shifts the peak toward higher energy. This behavior arises solely from the $\delta$-function, as the peak follows $E_\gamma = m_{\phi_3}$. The exclusion limits of the \texttt{COMPTEL} \cite{dissertation} and \texttt{EGRET} \cite{Strong:2004de} are shown by black and magenta shaded regions, respectively. It shows, the flux, for the prompt photon case, for all the $m_{\phi_3}$ considered, lies comfortably well below the exclusion limits, making the scenario viable and well motivated to be tested in future detectors.

The key distinction between the two approaches to gamma-ray spectrum analysis, cascaded decay and prompt decay, lies in the role of the mediator. In the cascaded decay scenario, the process does not involve $H$ mediation, and the photon spectrum is primarily determined by kinematics, as shown in Eqs.~[\ref{eq:photon-energy}--\ref{eq:photon-energy-spectra}].
In contrast, the prompt decay channel depends explicitly on $H$ mediation. Here, mediator suppression combined with the smallness of the mixing further reduces the annihilation cross section, and consequently the flux. 
As a result, the cascaded decay scenario is subject to stringent bounds, severely limiting predictivity, whereas the prompt decay case is more relaxed, making it particularly favorable for indirect detection prospects.
\subsection{Neutrino spectral features from cascaded annihilations}
\label{Netrino-spectrum}
The indirect detection of DM via neutrino signals provides a promising and complementary avenue, particularly in scenarios involving neutrinophilic mediators. The gauge structure of the model provides a dominant coupling of $H$ with a right-handed neutrino $N_R$ of mass 10 MeV, and the active neutrino $\nu_L$. Due to the light mass of the RHN, we do not consider further decay of the RHN. 
Therefore, we focus on the neutrino flux arising from $H\to N_R\, \nu_L$,  where the $H$ are pair-produced from $\phi_3$'s self-annihilation. Thus, the cascaded process results in a neutrino spectrum with a distinctive box-shaped characteristic~\cite{Ibarra:2012dw, Ibarra:2013eda, Boddy:2015efa}. In the lab frame, the energy depends on the angle $\theta$ between the direction of the outgoing neutrino and the boost direction of the intermediate particle $H$, given by,
\begin{align}\label{eq:Enu}
    E_{\nu} = \frac{m_{\phi_3}}{\sqrt{2}} \Big[1+\sqrt{1-\frac{m_H^2}{2m_{\phi_3}^2}} \sqrt{1-\frac{(m_{N_R}+m_{\nu_L})^2}{m_H^2}} \, \cos\theta \Big]\,,
\end{align}
where $\theta =0^\circ$ defines forward emission and $\theta =180^\circ$ corresponds to backward emission. The range of allowed neutrino energy is thus bounded between a maximum energy $E_\nu^{max}$ and a minimum energy $E_\nu^{min}$, respectively, corresponding to $\cos \theta= \pm 1$. The resulting spectrum is flat within this interval and is given by,
\begin{align} \label{eq:nu-spectrm}
    \frac{dN_{\nu}}{dE_{\nu}} = \frac{4}{\big(E_{\nu}^{max} - E_{\nu}^{min}\big)} \Theta\big(E_\nu - E_{\nu}^{min}\big) \, \Theta \big(E_{\nu}^{max} - E_{\nu}\big)\,,
\end{align}
where $\Theta(x)$ is the Heaviside step function ensuring the spectrum is non-zero only in the kinematically allowed range.
\begin{figure}[h!]
    \centering
    \includegraphics[width=0.5\linewidth]{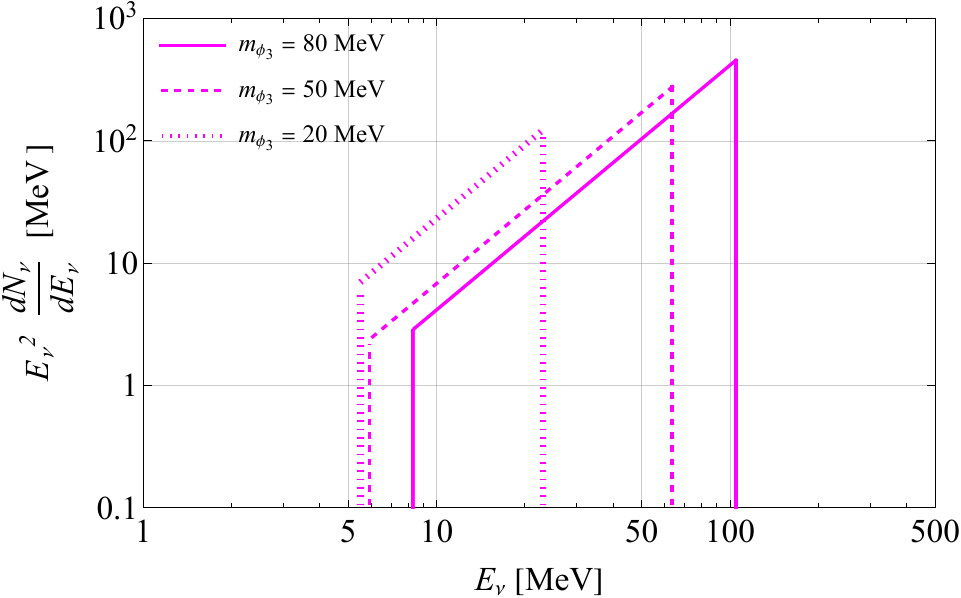}
    \caption{Variation of neutrino spectra $E_\nu^2 \, \frac{dN_\nu}{dE_\nu}$ with the outgoing neutrino energy, for various DM mass, with $H$'s mass fixed at $m_H=20$ MeV.}
    \label{fig:nu-spectrum}
\end{figure}
Fig. \ref{fig:nu-spectrum} shows the energy spectrum of the neutrinos for various DM masses with fixed $m_H=20$ MeV. The spectral shape for each curve exhibits a distinct box-like structure, in alignment with the analytical form given in Eq. \ref{eq:nu-spectrm}. The width of the box is again decided by the $\Delta E_{\nu}= E_{\nu}^{max} - E_{\nu}^{min}$, while $E_c=(E_{\nu}^{max} + E_{\nu}^{min})/2=m_{\phi_3}/\sqrt{2}$, is determined by the DM mass. Both the photon spectra and neutrino spectra behave similarly, since they abide overall the same kinematics as described in Sec.~\ref{spectrum-energy}. This behavior contrasts with scenarios where neutrinos or photons originate from cascade decays involving intermediate hadronic states \cite{Leane:2017vag} like pions \cite{Boddy:2015efa}. In such cases, the decay kinematics leads to broader and softer spectra, often lacking a sharp box-like feature. However, such hadronic decay channels are beyond the scope of the present study.

The flux of neutrinos coming out from the annihilation of DM at the galactic center is parametrized as~\cite{Akita:2025dhg},    
\begin{align}
    \frac{d\phi_\nu}{dE_\nu} = \frac{\langle \sigma v \rangle}{8 \pi m_{\phi_3}^2}\frac{1}{3} \frac{dN_\nu}{dE_\nu} \int_{\Delta \Omega} d\Omega \int_{l.o.s} ds \rho_{\phi_3}^2\,,
\end{align}
where $\langle \sigma v\rangle$ is the thermally averaged cross-section for the $\phi_3\,\phi_3\to H\,H$ process. The integral is performed along the l.o.s and over the solid angle $\Delta \Omega$ subtended by the observed region. The DM density is denoted by $\rho_{\phi_3}$. The flux is commonly recast in terms of the astrophysical factor $J_{ann}$ defined as,
\begin{align}
    \frac{d\phi_\nu}{dE_\nu} = \frac{\langle \sigma v \rangle}{8 \pi m_{\phi_3}^2}\frac{1}{3} \frac{dN_\nu}{dE_\nu} \frac{1}{\Delta \Omega}\int_{\Delta \Omega} d\Omega J_{ann}\,,    
\end{align}
where the $1/3$ denotes the average of the three flavors of neutrinos. In line with the gamma-ray analysis, we adopt a region of interest centered around the Galactic Center with $\Delta \Omega=1.3 \,~ \text{sr}$, and $\int_{\Delta \Omega} d\Omega J_{ann} = 9.2 \times 10^{22} ~\text{GeV}^2\,\text{cm}^{-5}\,\text{sr}$. This flux expression is used to compute the expected neutrino signal on Earth from DM annihilation at the GC. This will be convolved with detector energy resolution effects in the subsequent analysis. 
\begin{figure}[h!]
    \centering
    \includegraphics[scale=0.45]{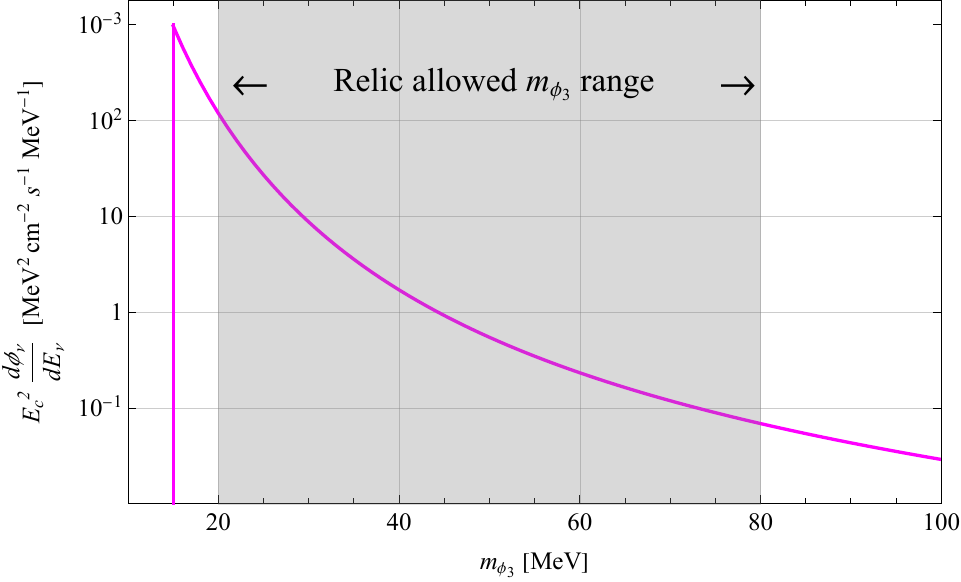} \quad \includegraphics[scale=0.45]{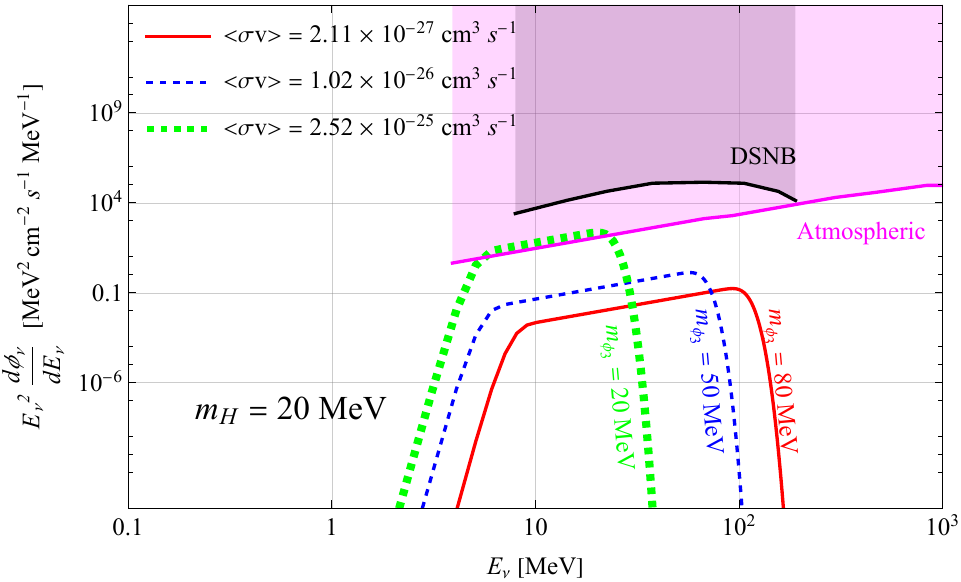}
    \caption{Left: Neutrino flux variation with the DM. The gray shaded region is the allowed relic density region. Right: The box-shaped neutrino spectrum shown for different mass choices of DM, by red solid line ($m_{\phi_3}=80$ MeV), blue dashed line ($m_{\phi_3}=50$ MeV), and green dotted line ($m_{\phi_3}=20$ MeV). For each case, the thermal averaged cross-section is mentioned in the inset. The \texttt{DSNB} and Atmospheric data make the upper boundary for the neutrino flux. The magenta and black shaded region is excluded.}
    \label{fig:nu-flux}
\end{figure}

Figure~\ref{fig:nu-flux} (left panel) shows the total neutrino flux (scaled by $E_c^2$) as a function of the DM mass $m_{\phi_3}$. The flux decreases with increasing DM mass, determined by the scaling of the annihilation cross-section $\sigma_{\phi_3 \phi_3 \to HH} \propto \frac{1}{m_{\phi_3}^2}$.
It is noteworthy that the available phase space for the final state particles is controlled by the factor $\sqrt{1-\frac{m_H^2}{2m_{\phi_3}^2}}$, which appears in the neutrino energy expression Eq.~\ref{eq:Enu}. This term becomes zero at the kinematic threshold $m_{\phi_3} = m_H/\sqrt{2} \simeq 14$ MeV, below which the annihilation process is forbidden. As a result, the flux sharply drops to zero at this point, setting an initial cut-off at $m_{\phi_3}\sim$ 14 MeV. The gray-shaded section represents the allowed mass range for the correct relic density, as defined earlier.
In the right panel, we illustrate the neutrino flux as a function of the neutrino energy $E_\nu$, for three representative values of the DM mass $m_{\phi_3} = $ 20 MeV (green dotted line), 50 MeV (blue dashed line), and 80 MeV (red solid line). As mentioned before, the cross-section increases for comparatively lighter DM, as reflected in the figure. The width and $E_c$ of the box are defined by the neutrino energy Eq.~\ref{eq:Enu}.
Overlaid on the figure are two important experimental upper limits for the MeV-scale neutrino energy spectrum: \texttt{DSNB} limit \cite{Vitagliano:2019yzm} as well as the bound from Atmospheric Neutrino flux measurements~\cite{Peres:2009xe}. These bounds define the exclusion regions in black (\texttt{DSNB}) and magenta (Atmospheric), respectively. The model predictions corresponding to $m_{\phi_3}=$ 80 MeV (solid red) and 50 MeV (dashed blue) lie comfortably below both bounds, suggesting that these parameter choices are phenomenologically viable and potentially detectable with improved neutrino flux sensitivity in next-generation experiments. On the other hand, the green dotted curve for $m_{\phi_3}=$ 20 MeV lies very close to the \texttt{DSNB} neutrino exclusion limit and overlaps with the sensitive region of the atmospheric neutrino. While not strictly excluded, this mass range is subject to tight experimental scrutiny, and any further increase in observational sensitivity, particularly in the $10 - 30$ MeV energy range, could begin to probe or constrain such scenarios. This highlights the complementarity between indirect neutrino detection and the model parameter space, with low mass DM candidates offering testable predictions near the threshold of experimental data. 
\subsection{Neutrino spectral feature due to prompt annihilation}
\label{2nu-final-state}
\begin{figure}[h!]
    \centering
    \includegraphics[scale=0.45]{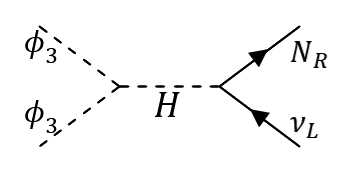}
    \caption{Feynamann diagram for $\phi_3 \,\phi_3 \to N_R\,\nu_L$, the prompt neutrino production from the self-annihilation of $\phi_3$ with $H$ mediation.}
    \label{fig:2nu}
\end{figure}

Beyond the cascaded decay channel discussed above, neutrinos can also be produced directly from the self-annihilation of $\phi_3$ following the $H$ mediated s-channel process as shown in Fig~\ref{fig:2nu}. Similar to the prompt gamma-ray spectrum, this channel produces a monochromatic neutrino line, which can be compared against the experimental bounds. The cross-section expression is given in \ref{eq:A3}.  

\begin{figure}[h!]
    \centering
    \includegraphics[scale=0.45]{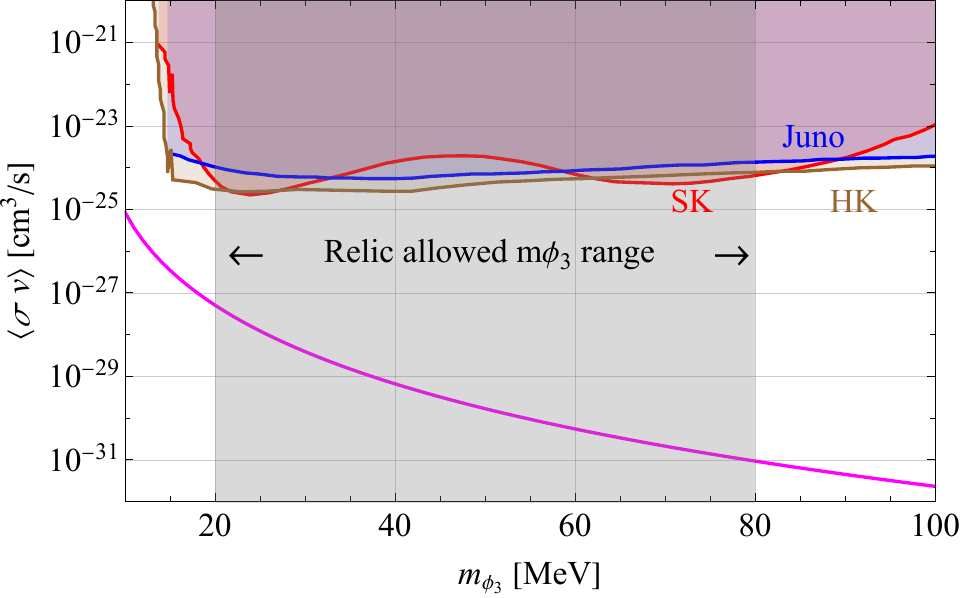} \quad \includegraphics[scale=0.45]{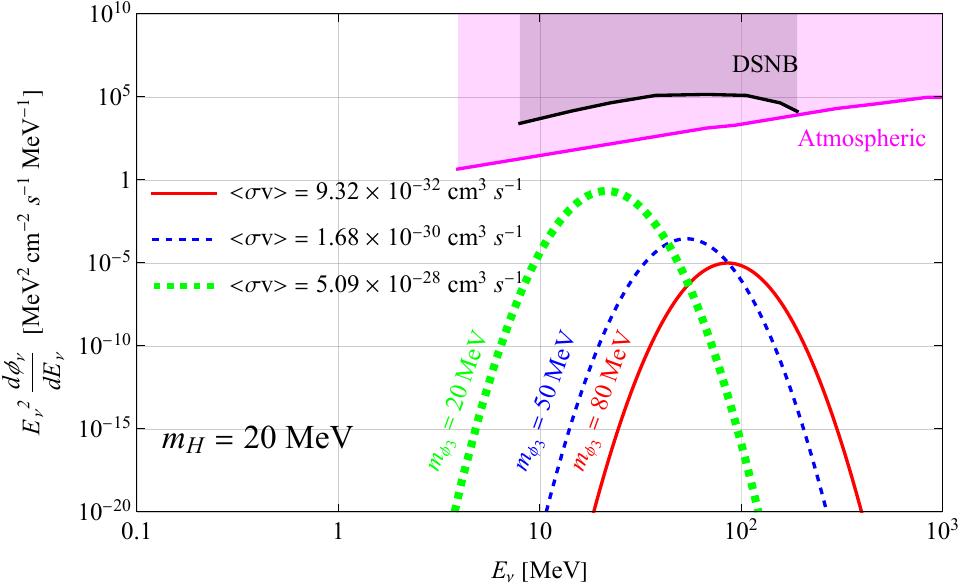}
    \caption{Left: 90\% CL exclusion limits on the thermal-averaged cross-section for the $\phi_3\,\phi_3 \to N_R \,\nu_L$ process in our galaxy from \texttt{SK} (red shaded), \texttt{JUNO} (blue shaded), and \texttt{HK} (brown shaded) sensitivity. The gray shaded region shows the allowed mass range for the correct relic density. Right: the flux of the neutrino spectrum generated due to the neutrino excess produced directly from the DM self-annihilation. The monochromatic neutrino flux is shown for the boosted scalar DM $\phi_3$ mass 80 MeV (red solid line), 50 MeV (blue dashed line), and 20 MeV (green dotted line). The bounds are from \texttt{DSNB}, and the atmospheric neutrino excess is shown by the black and magenta shaded regions.}
    \label{fig:2nu-spectra}
\end{figure}

In the left panel of Fig. \ref{fig:2nu-spectra}, the solid magenta line shows the thermal averaged cross-section for the $\phi_3\,\phi_3\to N_R\,\nu_L$, obtained for the model-dependent parameters. The two coupling constants $\lambda_{H\phi_3\phi_3}$ and $\lambda_{HN_R\nu_L}$ play an important role in controlling the cross-section. As discussed earlier, the coupling $\lambda_{H\phi_3\phi_3}$ is small due to mixing suppression, while the coupling constant $\lambda_{HN_R\nu_L}$ is suppressed due to the smallness of the Yukawa $y_\nu$, obtained from the analysis of the neutrino oscillation data \cite{Huitu:2017vye, PhysRevD.111.095016}. Therefore, along with the mediator suppression, the two coupling strengths play the decisive role to assure the magenta line lies well below the experimental bounds from the \texttt{SK} data \cite{Super-Kamiokande:2002hei} (red shaded region), \texttt{JUNO} \cite{JUNO:2023vyz} (blue shaded region), and the \texttt{HK} data \cite{Olivares-DelCampo:2018pdl} (brown shaded region). Once again, the gray shaded region defines the allowed mass range for the correct relic density.

It is interesting to note that, the neutrino spectrum from the self-annihilation of $\phi_3$, shows a delta function like shape similar to that of prompt gammay-ray scenario, as discussed in Sec. \ref{2gamma-final-state}. This spectrum is given by,
\begin{align}
    \frac{dN_\nu}{dE_\nu} = \delta(E_\nu - m_{\phi_3}).
\end{align}
Following the prescription as in Eq. \ref{eq:gamma-flux3}, together with the convolution function $\mathcal{G}(E,E_\nu)$ as defined in Eq. \ref{eq:Convolv-func}, we obtain the observed flux for the monochromatic neutrino. The results are shown by the red solid ($m_{\phi_3}=80$ MeV), blue dashed ($m_{\phi_3}=50$ MeV), and green dotted ($m_{\phi_3}=20$ MeV) lines in the right panel of the Fig. \ref{fig:2nu}. The exclusion limits from the \texttt{DSNB} \cite{Vitagliano:2019yzm} (black shaded region) and the Atmospheric neutrino excess \cite{Peres:2009xe} (magenta shaded region) lie safely well above the model-predicted peaks, keeping the scenario consistent with present data and providing a strong motivation for future searches.

\section{Conclusion}
\label{conclusion}
In this work, we explore the detection prospects of the two-component boosted light scalar DM. If the DM were not boosted, the resulting spectrum would be completely different. Since the boost depends on the mass hierarchy between two DM components, namely a light scalar and a heavier fermionic state, a scenario where the scalar DM has a mass comparable to that of the fermion DM, a negligible boost sets in. Therefore, a sizable boost requires the scalar DM to be much heavier, effectively pushing it beyond the MeV scale. However, the motivation behind studying the MeV-scale scalar DM lies in the persistent null results of WIMP searches at the GeV scale in both direct and indirect detection. Therefore, the non-boosted scalar DM scenario lies outside the scope of our present work.
Various theoretical and experimental bounds, along with the limit of correct relic density, provide the allowed mass range of the two DM candidates: $m_{\phi_3} \simeq 20~\text{MeV} - 80~\text{MeV}$, $m_{\chi_1} \simeq 30~\text{GeV} - 65~\text{GeV}$. Based on these mass ranges, we investigate $\phi_3$-nucleon/electron scattering. The cross section derived from the allowed parameter space of the model is shown in the $m_{\phi_3} - \sigma_{\phi_3\,n/e}$ plane as shown by Figs.~[\ref{fig:DM-N-sigma}, \ref{fig:DM-e-sigma}], together with exclusion limits from direct detection experiments such as \texttt{XENON10}, \texttt{XENON1T}, \texttt{DarkSide50}, and \texttt{SENSEI} experiments. It shows that the relic density allowed line lies beyond the sensitivity reach of the experiments. Moreover, when compared with the neutrino floor, the model-predicted cross section falls inside this region, placing the scenario in tension. Given the insufficient sensitivity from both the nuclear and electron recoil, the indirect detection prospect for this model is explored. 

The model is tested against both the gamma ray, as well as the neutrino spectrum. The annihilation of $\chi_1$ into $\phi_3$ makes the $\phi_3$ boosted, which subsequently annihilates into the BSM Higgs $H$. The Higgs then decays into photons or neutrinos. The resulting photon spectrum exhibits a characteristic box-like shape, with its position and width determined by the $\phi_3$ mass. We find that, for $m_{\phi_3}=$ 80 MeV and 50 MeV, the plateau region of the flux near the higher-energy end lies above the observational limits set by \texttt{COMPTEL} and \texttt{EGRET}, placing these mass ranges in tension with existing data. In contrast, the lower-energy side of the spectrum remains consistent with experimental data. However, lower masses such as $m_{\phi_3}=$ 20 MeV produce spectra that go above the exclusion limits, albeit challenging to resolve [cf. Fig.~\ref{fig:boxy}]. For the neutrino spectrum study,  our analysis shows that for DM masses $m_{\phi_3}=$ 80 MeV and 50 MeV, the predicted flux lies safely below the experimental bounds set by the \texttt{DSNB} and atmospheric neutrino data, while the 20 MeV case approaches the exclusion limits [cf. Fig.~\ref{fig:nu-flux}]. These results underline the importance of future low-energy neutrino observatories, which could detect or further constrain such box-like features, offering a complementary probe of the light dark sector in this model.  

In addition to the cascaded production of gamma-ray and neutrino spectra, the direct production of a pair of photons and neutrinos via the $H$ mediated annihilation of $\phi_3$ is also allowed. In this case, the cross-section explicitly depends on the decay strength of $H \to \gamma\gamma$ or $H \to N_R\,\nu_L$. The mediator suppression, along with the small couplings, assures the cross-section remains well below the exclusion limits from the \texttt{COMPTEL} and \texttt{EGRET} for gamma-ray analysis, and limits from the \texttt{SK}, \texttt{JUNO}, and \texttt{HK} (this is evident from Figs.~[\ref{fig:2gamma-spectra}, \ref{fig:2nu-spectra}]). In contrast to the cascaded process, here the energy spectrum primarily follows a $\delta$ function, leading to monochromatic lines at $E_{\nu/\gamma} = m_{\phi_3}$. After including detector resolution effects, the resulting flux remains below the present exclusion bounds. While the cascaded decay channels show that certain mass ranges are in mild tension with existing data, the prompt photon and neutrino channels remain entirely consistent and offer clear targets for future observations. The present analysis shows indirect detection as a powerful probe for MeV-scale boosted DM candidates, which are otherwise difficult to detect at usual direct search experiments.  

\section*{Acknowledgments}
We would like to thank Basabendu Barman and Amit Chakraborty for many useful discussions, careful reading of the manuscript, and insightful comments. We are also grateful to Dayana Joy, Shreecheta Chowdhury, Anjana Krishnan, and Rakesh Kumar Shivakumar for their careful reading and valuable comments. We also thank Dayana Joy for the system to work on. It is our pleasure to thank Yu Watanabe for valuable discussions during HPNP 2025 in Osaka, Japan.

\newpage
\appendix
\section{Dark Matter Annihilation cross sections}
\label{process}

\begin{itemize}
    \item {\fbox{$\phi_3\,\phi_3\to H\,H$}}

\begin{align} \label{eq:A1}\nonumber
    \sigma_{\phi_3\phi_3 \to HH} &= \frac{1}{16 \pi\, S} \frac{\sqrt{S-4m_H^2}}{\sqrt{S-4m_{\phi_3}^2}} \Big[ \lambda_{\phi_3\phi_3HH}^2+ \frac{(\lambda_{\phi_3\phi_3H} \, \lambda_{HHH})^2}{(S-m_H^2)^2} + \frac{8\, \lambda_{\phi_3\phi_3H}^4}{(4m_H^4-16m_H^2m_{\phi_3}^2+4Sm_{\phi_3}^2)} + \frac{2\, \lambda_{\phi_3\phi_3HH}\, \lambda_{\phi_3\phi_3H}\,\lambda_{HHH}}{(S-m_H^2)} \\ \nonumber 
    &+ \frac{8\,\lambda_{\phi_3\phi_3H}^4}{(4m_H^4-16m_H^2m_{\phi_3}^2+4Sm_{\phi_3}^2)} + \frac{4\, \lambda_{\phi_3\phi_3 H}^3 \, \lambda_{HHH}}{(S-4m_{\phi_3}^2)} \, \log \frac{2m_H^2-S+\sqrt{S-4m_H^2}\sqrt{S-4m_{\phi_3}^2}}{2m_H^2-S-\sqrt{S-4m_H^2}\sqrt{S-4m_{\phi_3}^2}} \\ 
   & + \frac{4\, \lambda_{\phi_3\phi_3HH}\,\lambda_{\phi_3\phi_3H}^2}{(S-4m_{\phi_3}^2)} \log \frac{2m_H^2-S+\sqrt{S-4m_H^2}\sqrt{S-4m_{\phi_3}^2}}{2m_H^2-S-\sqrt{S-4m_H^2}\sqrt{S-4m_{\phi_3}^2}}\Big]
\end{align}

\item {\fbox{$\phi_3\,\phi_3\to \gamma\,\gamma$}}
\begin{align}\label{eq:A2}
    \sigma_{\phi_3\phi_3\to\gamma\gamma} = \frac{1}{16 \pi\, S} \frac{1}{\sqrt{1-\frac{4m_{\phi_3}^2}{S}}} \frac{\lambda_{H\phi_3\phi_3}^2}{(S-m_H^2)^2+m_H^2(\Gamma_H^{tot})^2} 16\pi\, m_H\,\Gamma_{H\to \gamma\gamma}
\end{align}

\item {\fbox{$\phi_3\,\phi_3\to N_R\, \nu_L$}}
\begin{align}\label{eq:A3}
    \sigma_{\phi_3\phi_3\to N_R\nu_L} = \frac{1}{16 \pi\, S} \frac{\Big(1-\frac{(m_{N_R}+m_{\nu_L})^2}{S}\Big)^{3/2}}{\sqrt{1-\frac{4m_{\phi_3}^2}{S}}} \frac{\lambda_{H\phi_3\phi_3}^2 \lambda_{HN_R\nu_L}^2}{(S-m_H^2)^2+m_H^2(\Gamma_H^{tot})^2}
\end{align}

\end{itemize}

\section{Spectrum Energy for Cascaded process:}
\label{spectrum-energy}
$$
\phi_3\,\phi_3 \;\to\; H\,H \,, \qquad H \to x\,x \, .
$$
We work in the lab frame defined as the DM COM frame of the initial annihilation. Quantities with a star '$*$' are evaluated in the decaying particle's rest frame.

\subsection{Step 1: Production kinematics of $H$}
Let each produced $H$ carry energy $E_H$ and three-momentum magnitude $p_H$ in the lab/DM-CM frame. By definition
$$
p_H = \sqrt{E_H^2 - m_H^2}\,, \qquad 
\gamma \equiv \frac{E_H}{m_H}\,,\qquad 
\beta \equiv \frac{p_H}{E_H} = \sqrt{1-\frac{m_H^2}{E_H^2}}\,.
$$

\subsection{Step 2: Two-body decay $H\to x\,x$}
In the $H$ rest frame,
$$
E_x^* = \frac{m_H}{2}\,,\qquad 
p_x^* = \sqrt{(E_x^*)^2 - m_x^2} = \frac{1}{2}\sqrt{m_H^2 - 4 m_x^2}\,.
$$
If $\theta$ is the angle between the daughter's momentum and the boost direction of $H$ in the lab, then in the lab frame or the DM frame, the daughter energy  becomes,
\begin{equation}
E_x = \gamma\big( E_x^* + \beta\, p_x^* \cos\theta \big).
\end{equation}
Plugging in $E_x^*$, $p_x^*$, and $\gamma,\beta$:
\begin{align} \nonumber
E_x &= \frac{E_H}{m_H}\,\Big[\frac{m_H}{2} + \beta\,\frac{1}{2}\sqrt{m_H^2-4m_x^2}\,\cos\theta\Big] \\ 
&= \frac{E_H}{2}\,\Big[ 1 + \sqrt{1-\frac{m_H^2}{E_H^2}}\,\sqrt{1-\frac{4m_x^2}{m_H^2}}\,\cos\theta \Big]. \label{eq:Ex}
\end{align}
Using $E_H=\sqrt{2}\,m_{\phi_3}$, Eq.~\eqref{eq:Ex} becomes,
\begin{align}
  E_x = \frac{m_{\phi_3}}{\sqrt{2}}\,\Big[1+ \sqrt{1-\frac{m_H^2}{2m_{\phi_3}^2}}\,
\sqrt{1-\frac{4m_x^2}{m_H^2}}\,\cos\theta \Big].  
\end{align}
For gamma-ray spectrum, the $E_x$ becomes,
$$
E_{\gamma} = \frac{m_{\phi_3}}{\sqrt{2}} \Big( 1+ \cos \theta \sqrt{1-\frac{m_H^2}{2\,m_{\phi_3}^2}}~ \Big)\,,
$$
whereas for neutrino spectrum it is,
$$
E_{\nu} = \frac{m_{\phi_3}}{\sqrt{2}} \Big[1+\sqrt{1-\frac{m_H^2}{2m_{\phi_3}^2}} \sqrt{1-\frac{(m_{N_R}+m_{\nu_L})^2}{m_H^2}} \, \cos\theta \Big]\,.
$$


\begin{thebibliography}{10}

\bibitem{Planck:2018vyg}
N.~Aghanim et~al.
\newblock {Planck 2018 results. VI. Cosmological parameters}.
\newblock {\em Astron. Astrophys.}, 641:A6, 2020.
\newblock [Erratum: Astron.Astrophys. 652, C4 (2021)].

\bibitem{Bullock:1999he}
James~S. Bullock, Tsafrir~S. Kolatt, Yair Sigad, Rachel~S. Somerville, Andrey~V. Kravtsov, Anatoly~A. Klypin, Joel~R. Primack, and Avishai Dekel.
\newblock {Profiles of dark haloes. Evolution, scatter, and environment}.
\newblock {\em Mon. Not. Roy. Astron. Soc.}, 321:559--575, 2001.

\bibitem{Zwicky:1933gu}
F.~Zwicky.
\newblock {Die Rotverschiebung von extragalaktischen Nebeln}.
\newblock {\em Helv. Phys. Acta}, 6:110--127, 1933.

\bibitem{Begeman:1991iy}
K.~G. Begeman, A.~H. Broeils, and R.~H. Sanders.
\newblock {Extended rotation curves of spiral galaxies: Dark haloes and modified dynamics}.
\newblock {\em Mon. Not. Roy. Astron. Soc.}, 249:523, 1991.

\bibitem{Bertone:2010zza}
J.~Silk et~al.
\newblock {\em {Particle Dark Matter: Observations, Models and Searches}}.
\newblock Cambridge Univ. Press, Cambridge, 2010.

\bibitem{Bauer:2017qwy}
Martin Bauer and Tilman Plehn.
\newblock {\em {Yet Another Introduction to Dark Matter}: {The Particle Physics Approach}}, volume 959 of {\em Lecture Notes in Physics}.
\newblock Springer, 2019.

\bibitem{Bertone:2004pz}
Gianfranco Bertone, Dan Hooper, and Joseph Silk.
\newblock {Particle dark matter: Evidence, candidates and constraints}.
\newblock {\em Phys. Rept.}, 405:279--390, 2005.

\bibitem{Lisanti:2016jxe}
Mariangela Lisanti.
\newblock {Lectures on Dark Matter Physics}.
\newblock In {\em {Theoretical Advanced Study Institute in Elementary Particle Physics}: {New Frontiers in Fields and Strings}}, pages 399--446, 2017.

\bibitem{refId0}
N.~Aghanim et~al.
\newblock {Planck 2018 results. VI. Cosmological parameters}.
\newblock {\em Astron. Astrophys.}, 641:A6, 2020.
\newblock [Erratum: Astron.Astrophys. 652, C4 (2021)].

\bibitem{Arcadi:2017kky}
Giorgio Arcadi, Ma\'\i{}ra Dutra, Pradipta Ghosh, Manfred Lindner, Yann Mambrini, Mathias Pierre, Stefano Profumo, and Farinaldo~S. Queiroz.
\newblock {The waning of the WIMP? A review of models, searches, and constraints}.
\newblock {\em Eur. Phys. J. C}, 78(3):203, 2018.

\bibitem{Planck:2015fie}
P.~A.~R. Ade et~al.
\newblock {Planck 2015 results. XIII. Cosmological parameters}.
\newblock {\em Astron. Astrophys.}, 594:A13, 2016.

\bibitem{Arcadi:2024ukq}
Giorgio Arcadi, David Cabo-Almeida, Ma{\'\i}ra Dutra, Pradipta Ghosh, Manfred Lindner, Yann Mambrini, Jacinto~P. Neto, Mathias Pierre, Stefano Profumo, and Farinaldo~S. Queiroz.
\newblock {The Waning of the WIMP: Endgame?}
\newblock {\em Eur. Phys. J. C}, 85(2):152, 2025.

\bibitem{XENON:2023cxc}
E.~Aprile et~al.
\newblock {First Dark Matter Search with Nuclear Recoils from the XENONnT Experiment}.
\newblock {\em Phys. Rev. Lett.}, 131(4):041003, 2023.

\bibitem{LZ:2022lsv}
J.~Aalbers et~al.
\newblock {First Dark Matter Search Results from the LUX-ZEPLIN (LZ) Experiment}.
\newblock {\em Phys. Rev. Lett.}, 131(4):041002, 2023.

\bibitem{Bringmann:2018cvk}
Torsten Bringmann and Maxim Pospelov.
\newblock {Novel direct detection constraints on light dark matter}.
\newblock {\em Phys. Rev. Lett.}, 122(17):171801, 2019.

\bibitem{Das:2021lcr}
Anirban Das and Manibrata Sen.
\newblock {Boosted dark matter from diffuse supernova neutrinos}.
\newblock {\em Phys. Rev. D}, 104(7):075029, 2021.

\bibitem{Xia:2022tid}
Chen Xia, Yan-Hao Xu, and Yu-Feng Zhou.
\newblock {Azimuthal asymmetry in cosmic-ray boosted dark matter flux}.
\newblock {\em Phys. Rev. D}, 107(5):055012, 2023.

\bibitem{Jho:2021rmn}
Yongsoo Jho, Jong-Chul Park, Seong~Chan Park, and Po-Yan Tseng.
\newblock {Cosmic-Neutrino-Boosted Dark Matter ($\nu$BDM)}.
\newblock 1 2021.

\bibitem{Yin:2018yjn}
Wen Yin.
\newblock {Highly-boosted dark matter and cutoff for cosmic-ray neutrinos through neutrino portal}.
\newblock {\em EPJ Web Conf.}, 208:04003, 2019.

\bibitem{Herrera:2023fpq}
Gonzalo Herrera, Alejandro Ibarra, and Satoshi Shirai.
\newblock {Enhanced prospects for direct detection of inelastic dark matter from a non-galactic diffuse component}.
\newblock {\em JCAP}, 04:026, 2023.

\bibitem{Herrera:2021puj}
Gonzalo Herrera and Alejandro Ibarra.
\newblock {Direct detection of non-galactic light dark matter}.
\newblock {\em Phys. Lett. B}, 820:136551, 2021.

\bibitem{Bhowmick:2022zkj}
Supritha Bhowmick, Diptimoy Ghosh, and Divya Sachdeva.
\newblock {Blazar boosted dark matter \textemdash{} direct detection constraints on \ensuremath{\sigma}e\ensuremath{\chi}: role of energy dependent cross sections}.
\newblock {\em JCAP}, 07:039, 2023.

\bibitem{Wang:2021jic}
Jin-Wei Wang, Alessandro Granelli, and Piero Ullio.
\newblock {Direct Detection Constraints on Blazar-Boosted Dark Matter}.
\newblock {\em Phys. Rev. Lett.}, 128(22):221104, 2022.

\bibitem{Granelli:2022ysi}
Alessandro Granelli, Piero Ullio, and Jin-Wei Wang.
\newblock {Blazar-boosted dark matter at Super-Kamiokande}.
\newblock {\em JCAP}, 07(07):013, 2022.

\bibitem{Maity:2022exk}
Tarak~Nath Maity and Ranjan Laha.
\newblock {Cosmic-ray boosted dark matter in Xe-based direct detection experiments}.
\newblock {\em Eur. Phys. J. C}, 84(2):117, 2024.

\bibitem{Agashe:2015xkj}
Kaustubh Agashe, Yanou Cui, Lina Necib, and Jesse Thaler.
\newblock {(In)Direct Detection of Boosted Dark Matter}.
\newblock {\em J. Phys. Conf. Ser.}, 718(4):042041, 2016.

\bibitem{Aoki:2018gjf}
Mayumi Aoki and Takashi Toma.
\newblock {Boosted Self-interacting Dark Matter in a Multi-component Dark Matter Model}.
\newblock {\em JCAP}, 10:020, 2018.

\bibitem{Nagao:2024itk}
Keiko~I. Nagao.
\newblock {Direct detection of boosted dark matter in two component dark matter scenario}.
\newblock In {\em {29th International Symposium on Particles, String and Cosmology}}, 12 2024.

\bibitem{PhysRevD.111.095016}
Arindam Basu, Amit Chakraborty, Nilanjana Kumar, and Soumya Sadhukhan.
\newblock Viability of boosted light dark matter in a two-component scenario.
\newblock {\em Phys. Rev. D}, 111:095016, May 2025.

\bibitem{Grimus:2007if}
W.~Grimus, L.~Lavoura, O.~M. Ogreid, and P.~Osland.
\newblock {A Precision constraint on multi-Higgs-doublet models}.
\newblock {\em J. Phys. G}, 35:075001, 2008.

\bibitem{Grimus:2008nb}
W.~Grimus, L.~Lavoura, O.~M. Ogreid, and P.~Osland.
\newblock {The Oblique parameters in multi-Higgs-doublet models}.
\newblock {\em Nucl. Phys. B}, 801:81--96, 2008.

\bibitem{Haber:2010bw}
Howard~E. Haber and Deva O'Neil.
\newblock {Basis-independent methods for the two-Higgs-doublet model III: The CP-conserving limit, custodial symmetry, and the oblique parameters S, T, U}.
\newblock {\em Phys. Rev. D}, 83:055017, 2011.

\bibitem{XENON10:2011prx}
J.~Angle et~al.
\newblock {A search for light dark matter in XENON10 data}.
\newblock {\em Phys. Rev. Lett.}, 107:051301, 2011.
\newblock [Erratum: Phys.Rev.Lett. 110, 249901 (2013)].

\bibitem{XENON:2019gfn}
E.~Aprile et~al.
\newblock {Light Dark Matter Search with Ionization Signals in XENON1T}.
\newblock {\em Phys. Rev. Lett.}, 123(25):251801, 2019.

\bibitem{DarkSide:2022dhx}
P.~Agnes et~al.
\newblock {Search for Dark-Matter{\textendash}Nucleon Interactions via Migdal Effect with DarkSide-50}.
\newblock {\em Phys. Rev. Lett.}, 130(10):101001, 2023.

\bibitem{SENSEI:2023zdf}
Prakruth Adari et~al.
\newblock {First Direct-Detection Results on Sub-GeV Dark Matter Using the SENSEI Detector at SNOLAB}.
\newblock {\em Phys. Rev. Lett.}, 134(1):011804, 2025.

\bibitem{Photino.55.257}
Joseph Silk, Keith Olive, and Mark Srednicki.
\newblock The photino, the sun, and high-energy neutrinos.
\newblock {\em Phys. Rev. Lett.}, 55:257--259, Jul 1985.

\bibitem{Ullio:2002pj}
Piero Ullio, Lars Bergstrom, Joakim Edsjo, and Cedric~G. Lacey.
\newblock {Cosmological dark matter annihilations into gamma-rays - a closer look}.
\newblock {\em Phys. Rev. D}, 66:123502, 2002.

\bibitem{dissertation}
Georg Weidenspointner.
\newblock {\em The Origin of the Cosmic Gamma-Ray Background in the COMPTEL Energy Range}.
\newblock PhD thesis, Technische Universität München, 1999.

\bibitem{Strong:2004de}
Andrew~W. Strong, Igor~V. Moskalenko, and Olaf Reimer.
\newblock {Diffuse galactic continuum gamma rays. A Model compatible with EGRET data and cosmic-ray measurements}.
\newblock {\em Astrophys. J.}, 613:962--976, 2004.

\bibitem{Vitagliano:2019yzm}
Edoardo Vitagliano, Irene Tamborra, and Georg Raffelt.
\newblock {Grand Unified Neutrino Spectrum at Earth: Sources and Spectral Components}.
\newblock {\em Rev. Mod. Phys.}, 92:45006, 2020.

\bibitem{Peres:2009xe}
Orlando L.~G. Peres and A.~Yu. Smirnov.
\newblock {Oscillations of very low energy atmospheric neutrinos}.
\newblock {\em Phys. Rev. D}, 79:113002, 2009.

\bibitem{Super-Kamiokande:2002hei}
M.~Malek et~al.
\newblock {Search for supernova relic neutrinos at SUPER-KAMIOKANDE}.
\newblock {\em Phys. Rev. Lett.}, 90:061101, 2003.

\bibitem{JUNO:2023vyz}
Angel Abusleme et~al.
\newblock {JUNO sensitivity to the annihilation of MeV dark matter in the galactic halo}.
\newblock {\em JCAP}, 09:001, 2023.

\bibitem{Olivares-DelCampo:2018pdl}
Andres Olivares-Del~Campo, Sergio Palomares-Ruiz, and Silvia Pascoli.
\newblock {Implications of a Dark Matter-Neutrino Coupling at Hyper-Kamiokande}.
\newblock In {\em {53rd Rencontres de Moriond on Electroweak Interactions and Unified Theories}}, pages 441--444, 2018.

\bibitem{Machado:2015sha}
P.~A.~N. Machado, Y.~F. Perez, O.~Sumensari, Z.~Tabrizi, and R.~Zukanovich Funchal.
\newblock {On the Viability of Minimal Neutrinophilic Two-Higgs-Doublet Models}.
\newblock {\em JHEP}, 12:160, 2015.

\bibitem{Nomura:2017jxb}
Takaaki Nomura and Hiroshi Okada.
\newblock {Neutrinophilic two Higgs doublet model with dark matter under an alternative $U(1)_{B-L}$ gauge symmetry}.
\newblock {\em Eur. Phys. J. C}, 78(3):189, 2018.

\bibitem{Witten}
Mark~W. Goodman and Edward Witten.
\newblock Detectability of certain dark-matter candidates.
\newblock {\em Phys. Rev. D}, 31:3059--3063, Jun 1985.

\bibitem{LUX:2016ggv}
D.~S. Akerib et~al.
\newblock {Results from a search for dark matter in the complete LUX exposure}.
\newblock {\em Phys. Rev. Lett.}, 118(2):021303, 2017.

\bibitem{PandaX-II:2017hlx}
Xiangyi Cui et~al.
\newblock {Dark Matter Results From 54-Ton-Day Exposure of PandaX-II Experiment}.
\newblock {\em Phys. Rev. Lett.}, 119(18):181302, 2017.

\bibitem{XENON:2018voc}
E.~Aprile et~al.
\newblock {Dark Matter Search Results from a One Ton-Year Exposure of XENON1T}.
\newblock {\em Phys. Rev. Lett.}, 121(11):111302, 2018.

\bibitem{Ema:2020ulo}
Yohei Ema, Filippo Sala, and Ryosuke Sato.
\newblock {Neutrino experiments probe hadrophilic light dark matter}.
\newblock {\em SciPost Phys.}, 10(3):072, 2021.

\bibitem{Weisenpacher:2000ip}
Peter Weisenpacher.
\newblock {Origin of the nucleon electromagnetic form-factors dipole formula}.
\newblock {\em Czech. J. Phys.}, 51:785--790, 2001.

\bibitem{PhysRevLett.115.092301}
Martin Hoferichter, Jacobo Ruiz~de Elvira, Bastian Kubis, and Ulf-G. Mei\ss{}ner.
\newblock High-precision determination of the pion-nucleon $\ensuremath{\sigma}$ term from roy-steiner equations.
\newblock {\em Phys. Rev. Lett.}, 115:092301, Aug 2015.

\bibitem{Essig:2019xkx}
Rouven Essig, Josef Pradler, Mukul Sholapurkar, and Tien-Tien Yu.
\newblock {Relation between the Migdal Effect and Dark Matter-Electron Scattering in Isolated Atoms and Semiconductors}.
\newblock {\em Phys. Rev. Lett.}, 124(2):021801, 2020.

\bibitem{Herrera:2023xun}
Gonzalo Herrera.
\newblock {A neutrino floor for the Migdal effect}.
\newblock {\em JHEP}, 05:288, 2024.

\bibitem{Duda:2006uk}
Gintaras Duda, Ann Kemper, and Paolo Gondolo.
\newblock {Model Independent Form Factors for Spin Independent Neutralino-Nucleon Scattering from Elastic Electron Scattering Data}.
\newblock {\em JCAP}, 04:012, 2007.

\bibitem{Helm:1956}
Richard~H. Helm.
\newblock Inelastic and elastic scattering of 187-mev electrons from selected even-even nuclei.
\newblock {\em Phys. Rev.}, 104:1466--1475, Dec 1956.

\bibitem{Lin:2019uvt}
Tongyan Lin.
\newblock {Dark matter models and direct detection}.
\newblock {\em PoS}, 333:009, 2019.

\bibitem{Barman:2024lxy}
Basabendu Barman, Arindam Das, and Sanjoy Mandal.
\newblock {Dark matter-electron scattering and freeze-in scenarios in the light of Z' mediation}.
\newblock {\em Phys. Rev. D}, 110(5):055029, 2024.

\bibitem{DarkSide:2018bpj}
P.~Agnes et~al.
\newblock {Low-Mass Dark Matter Search with the DarkSide-50 Experiment}.
\newblock {\em Phys. Rev. Lett.}, 121(8):081307, 2018.

\bibitem{PhysRevD.109.083016}
Ben Carew, Ashlee~R. Caddell, Tarak~Nath Maity, and Ciaran A.~J. O'Hare.
\newblock Neutrino fog for dark matter-electron scattering experiments.
\newblock {\em Phys. Rev. D}, 109:083016, Apr 2024.

\bibitem{Trickle:2020oki}
Tanner Trickle, Zhengkang Zhang, and Kathryn~M. Zurek.
\newblock {Effective field theory of dark matter direct detection with collective excitations}.
\newblock {\em Phys. Rev. D}, 105(1):015001, 2022.

\bibitem{Gu:2017gle}
Pei-Hong Gu and Xiao-Gang He.
\newblock {Electrophilic dark matter with dark photon: from DAMPE to direct detection}.
\newblock {\em Phys. Lett. B}, 778:292--295, 2018.

\bibitem{Essig:2011nj}
Rouven Essig, Jeremy Mardon, and Tomer Volansky.
\newblock {Direct Detection of Sub-GeV Dark Matter}.
\newblock {\em Phys. Rev. D}, 85:076007, 2012.

\bibitem{Nagao:2024hit}
Keiko~I. Nagao, Tatsuhiro Naka, and Takaaki Nomura.
\newblock {Directional direct detection of MeV scale boosted dark matter in two component dark matter scenario via dark photon interaction}.
\newblock {\em JCAP}, 04:030, 2025.

\bibitem{OHare:2015utx}
Ciaran A.~J. O'Hare, Anne~M. Green, Julien Billard, Enectali Figueroa-Feliciano, and Louis~E. Strigari.
\newblock {Readout strategies for directional dark matter detection beyond the neutrino background}.
\newblock {\em Phys. Rev. D}, 92(6):063518, 2015.

\bibitem{Bergstrom:1997fj}
Lars Bergstrom, Piero Ullio, and James~H. Buckley.
\newblock {Observability of gamma-rays from dark matter neutralino annihilations in the Milky Way halo}.
\newblock {\em Astropart. Phys.}, 9:137--162, 1998.

\bibitem{Coogan:2019uij}
Adam Coogan, Benjamin~V. Lehmann, and Stefano Profumo.
\newblock {Connecting direct and indirect detection with a dark spike in the cosmic-ray electron spectrum}.
\newblock {\em JCAP}, 10:063, 2019.

\bibitem{Arina:2025ner}
Chiara Arina, Mattia Di~Mauro, Nicolao Fornengo, Jan Heisig, Adil Jueid, and Roberto~Ruiz de~Austri.
\newblock {CosmiXs: Improved spectra for dark matter indirect detection}.
\newblock 1 2025.

\bibitem{Ibarra:2012dw}
Alejandro Ibarra, Sergio Lopez~Gehler, and Miguel Pato.
\newblock {Dark matter constraints from box-shaped gamma-ray features}.
\newblock {\em JCAP}, 07:043, 2012.

\bibitem{Ibarra:2013eda}
Alejandro Ibarra, Hyun~Min Lee, Sergio L{\'o}pez~Gehler, Wan-Il Park, and Miguel Pato.
\newblock {Gamma-ray boxes from axion-mediated dark matter}.
\newblock {\em JCAP}, 05:016, 2013.
\newblock [Erratum: JCAP 03, E01 (2016)].

\bibitem{Boddy:2015efa}
Kimberly~K. Boddy and Jason Kumar.
\newblock {Indirect Detection of Dark Matter Using MeV-Range Gamma-Ray Telescopes}.
\newblock {\em Phys. Rev. D}, 92(2):023533, 2015.

\bibitem{Navarro:1995iw}
Julio~F. Navarro, Carlos~S. Frenk, and Simon D.~M. White.
\newblock {The Structure of cold dark matter halos}.
\newblock {\em Astrophys. J.}, 462:563--575, 1996.

\bibitem{Navarro:1996gj}
Julio~F. Navarro, Carlos~S. Frenk, and Simon D.~M. White.
\newblock {A Universal density profile from hierarchical clustering}.
\newblock {\em Astrophys. J.}, 490:493--508, 1997.

\bibitem{Vertongen:2011mu}
Gilles Vertongen and Christoph Weniger.
\newblock {Hunting Dark Matter Gamma-Ray Lines with the Fermi LAT}.
\newblock {\em JCAP}, 05:027, 2011.

\bibitem{Bartels:2017dpb}
Richard Bartels, Daniele Gaggero, and Christoph Weniger.
\newblock {Prospects for indirect dark matter searches with MeV photons}.
\newblock {\em JCAP}, 05:001, 2017.

\bibitem{Cirelli:2025rky}
Marco Cirelli, Arpan Kar, and Halim Shaikh.
\newblock {Indirect searches for realistic sub-GeV Dark Matter models}.
\newblock 8 2025.

\bibitem{Guo:2023kqt}
Jun Guo, Lei Wu, and Bin Zhu.
\newblock {MeV gamma-ray constraints for light dark matter from semi-annihilation}.
\newblock {\em Phys. Lett. B}, 840:137853, 2023.

\bibitem{Djouadi:2005gi}
Abdelhak Djouadi.
\newblock {The Anatomy of electro-weak symmetry breaking. I: The Higgs boson in the standard model}.
\newblock {\em Phys. Rept.}, 457:1--216, 2008.

\bibitem{Palomares-Ruiz:2007egs}
Sergio Palomares-Ruiz.
\newblock {Model-independent bound on the dark matter lifetime}.
\newblock {\em Phys. Lett. B}, 665:50--53, 2008.

\bibitem{Akita:2025dhg}
Kensuke Akita, Alejandro Ibarra, and Robert Zimmermann.
\newblock {Dark matter explanations for the neutrino emission from the Seyfert galaxy NGC 1068}.
\newblock 7 2025.

\bibitem{Leane:2017vag}
Rebecca~K. Leane, Kenny C.~Y. Ng, and John~F. Beacom.
\newblock {Powerful Solar Signatures of Long-Lived Dark Mediators}.
\newblock {\em Phys. Rev. D}, 95(12):123016, 2017.

\bibitem{Huitu:2017vye}
Katri Huitu, Timo~J. K\"arkk\"ainen, Subhadeep Mondal, and Santosh~Kumar Rai.
\newblock {Exploring collider aspects of a neutrinophilic Higgs doublet model in multilepton channels}.
\newblock {\em Phys. Rev. D}, 97(3):035026, 2018.

\end{thebibliography}

\end{document}